\RequirePackage{lineno}

\documentclass[preprint,onecolumn,aps,prd,groupedaddress,nofootinbib,showpacs]{revtex4}
\usepackage{graphicx}
\graphicspath{{figures/}} \DeclareGraphicsExtensions{.eps}
\usepackage{amsmath}
\usepackage{bm}
\usepackage{slashed}
\usepackage[hypertex]{hyperref}
\usepackage{xcolor}
\usepackage{amsfonts}
\usepackage{soul}

\allowdisplaybreaks

\newcommand\gev{\mathrm{~GeV}}

\newcommand\msbar{\overline{\mathrm{MS}}}

\newcommand{\jpsi}{{J/\psi}}
\newcommand\psip{{\psi^\prime}}

\newcommand{\state}[4]{{^#1\hspace{-0.6mm}#2_{#3}^{[#4]}}}
\newcommand{\statenc}[3]{{^#1\hspace{-0.6mm}#2_{#3}}}
\newcommand\CSaSz{\state{1}{S}{0}{1}}
\newcommand\CSaPa{\state{1}{P}{1}{1}}
\newcommand\CScSa{\state{3}{S}{1}{1}}
\newcommand\CScPz{\state{3}{P}{0}{1}}
\newcommand\CScPa{\state{3}{P}{1}{1}}
\newcommand\CScPb{\state{3}{P}{2}{1}}
\newcommand\CScPj{\state{3}{P}{J}{1}}
\newcommand\COaSz{\state{1}{S}{0}{8}}
\newcommand\COaPa{\state{1}{P}{1}{8}}
\newcommand\COcSa{\state{3}{S}{1}{8}}

\newcommand\COcPj{\state{3}{P}{J}{8}}

\newcommand{\initialstate}[2]{{#1^{[#2]}}}
\newcommand\vone{\initialstate{v}{1}}
\newcommand\veight{\initialstate{v}{8}}
\newcommand\aone{\initialstate{a}{1}}
\newcommand\aeight{\initialstate{a}{8}}
\newcommand\tone{\initialstate{t}{1}}
\newcommand\teight{\initialstate{t}{8}}

\newcommand\DeltaZero{\Delta_0}

\newcommand\DPlusOne{\Delta_+^{[1]}}

\newcommand\DPlusEight{\Delta_-^{[8]}}

\newcommand\DMinusOne{\Delta_-^{[1]}}

\newcommand\DMinusEight{\Delta_+^{[8]}}

\newcommand\DPMOne{\Delta_\pm^{[1]}}
\newcommand\DPMEight{\Delta_\pm^{[8]}}

\newcommand\DAA{{\delta(1-z+\zeta_1)}}
\newcommand\DBB{{\delta(1-z-\zeta_1)}}
\newcommand\DXX{{\delta(1-z+\zeta_2)}}
\newcommand\DYY{{\delta(1-z-\zeta_2)}}

\newcommand\as{\alpha_s}

\newcommand\LogUV{\,\text{ln}\big[\frac{\mu_0^2}{m_Q^2}\big]}

\newcommand{\mylog}[1]{{\,\text{ln}(#1)}}
\newcommand\logtwo{\,\text{ln}\, 2}

\newcommand{\Plusz}[1]{{\left( #1 \right)_{0+}}}

\newcommand\Plusa{\left(\frac{1}{\zeta_1}\right)_{1+}}
\newcommand\Plusb{\left(\frac{1}{\zeta_1^2}\right)_{2+}}

\newcommand\PlusLoga{\left(\frac{\text{ln}(\zeta_1^2)}{\zeta_1}\right)_{1+}}

\newcommand{\cc}{{Q\bar{Q}}}

\newcommand{\ben}{\begin{eqnarray}}
\newcommand{\een}{\end{eqnarray}}

\newcommand{\bef}{\begin{figure}[!htp]}
\newcommand{\eef}{\end{figure}}

\newcommand{\bea}{\begin{eqnarray}}
\newcommand{\eea}{\end{eqnarray}}

\newcommand{\Pwave}{Pwave}

\begin{document}

\title{Heavy quarkonium fragmentation functions from a heavy quark pair. I. $S$ wave}

\author{Yan-Qing Ma}
\email{yqma@bnl.gov}
\affiliation{Physics Department, Brookhaven
National Laboratory, Upton, NY 11973, USA}

\author{Jian-Wei Qiu}
\email{jqiu@bnl.gov}
\affiliation{Physics Department, Brookhaven
National Laboratory, Upton, NY 11973, USA}
\affiliation{C.N. Yang Institute for Theoretical Physics and Department of Physics and Astronomy,
Stony Brook University, Stony Brook, NY 11794, USA}

\author{Hong Zhang}
\email{hong.zhang@stonybrook.edu}
\affiliation{Department of Physics and Astronomy, Stony Brook University, Stony Brook, NY 11794, USA}

\date{\today}

\preprint{YITP-SB-13-49}
\begin{abstract}

A QCD factorization formalism was recently proposed for evaluating heavy quarkonium production at large $p_T$ at collider energies. With systematically calculated short-distance partonic hard parts and evolution kernels of fragmentation functions (FFs), the predictive power of this factorization approach relies on our knowledge of a large number of universal FFs at an input factorization scale $\mu_0\gtrsim 2m_Q$ with heavy quark mass $m_Q$.  With the large heavy quark mass, the relative motion of the heavy quark pair inside a heavy quarkonium is effectively non-relativistic.  We evaluate these universal input FFs using non-relativistic QCD (NRQCD) factorization, and express the large number of FFs in terms of a few universal NRQCD long-distance matrix elements (LDMEs) with perturbatively calculated coefficients.  We derive complete contributions to the single-parton FFs at both ${\cal O}(\alpha_s)$ and ${\cal O}(\alpha_s^2)$, and the heavy quark pair FFs at ${\cal O}(\alpha_s)$.  We present detailed derivation for all contributions involving LDMEs of $S$-wave NRQCD $\cc$-states ($P$-wave contributions in a companion paper \cite{Pwave}).   Our results bridge the gap between the QCD factorization formalism and its phenomenological applications.
\end{abstract}


\maketitle


\section{introduction}

Heavy quarkonium production has been a powerful tool to test and challenge our understanding of strong interaction and QCD \cite{Brambilla:2010cs,Bodwin:2013nua}.  Since the heavy quark mass, $m_Q\gg\Lambda_{\rm QCD}$, the production of the heavy quark pair could be calculated perturbatively \cite{Collins:1985gm}. However, the transformation or hadronization of the pair to a heavy quarkonium is intrinsically nonperturbative. Different treatments of the hadronization process lead to various factorization models for heavy quarkonium production, such as color singlet model (CSM), color evaporation model (CEM) and non-relativistic QCD (NRQCD) model \cite{Brambilla:2010cs}. Among them, NRQCD model is by far the most successful in phenomenological study \cite{Brambilla:2010cs, Kramer:2001hh, Braaten:1996pv, Petrelli:1997ge}.

NRQCD model  \cite{Bodwin:1994jh}, which includes CSM and CEM as its special cases, is basically an effective field theory approach relied on the separation of momentum scales in heavy quarkonium production.  As a conjecture, NRQCD model separates heavy quarkonium production into two steps: (1) the production of a heavy quark pair of a particular spin and color state in a hard collision with a momentum transfer larger than twice of the heavy quark mass, which could be calculated perturbatively; and (2) the heavy quark pair then evolves into a physical heavy quarkonium, which is characterized by momentum scales much less than the heavy quark mass and is in principle a nonperturbative process, and the net transition rate is given by universal NRQCD long-distance matrix elements (LDMEs). Summing over the pair's all possible spin and color states gives the total inclusive cross section.  With the perturbative hard parts calculated to next-to-leading order (NLO) in $\as$ and carefully fitted LDMEs, NRQCD is very successful in interpreting the data on production rate of $\chi_{cJ}$, $J/\psi$ and $\Upsilon$ from Tevatron and the LHC \cite{Ma:2010vd,Ma:2010yw,Butenschoen:2010rq,Wang:2012is,Gong:2013qka}.

However, with additional large scales other than the heavy quark mass, potentially, the perturbative expansion of the hard parts of NRQCD factorization approach could be unstable. For example, for heavy quarkonium produced at large transverse momentum $p_T$, large $\mylog{p_T^2/m_Q^2}$-type logarithms need to be systematically resumed. Moreover, high order corrections may receive huge power enhancements in terms of ${p_T^2}/{m_Q^2}$, which could overwhelm the suppression of $\as$ at large $p_T$ \cite{KMQS-hq1,KMQS-hq2}.

Several inconsistencies between NLO NRQCD calculations and experimental data have been realized recently. The combination of color octet LDMEs for $\jpsi$ production, $M_{0,3.9}^{J/\psi}=7.4\times 10^{-2}$~GeV$^3$ \cite{Ma:2010yw}, obtained by fitting hadron collider data based on NLO NRQCD calculation, contradicts to the upper limit, $2.0 \times 10^{-2} \gev^3$, derived from $e^+ e^-$ data \cite{Zhang:2009ym}. The first attempt of global fitting on $J/\psi$ production in Ref.~\cite{Butenschoen:2010rq} effectively confirmed this inconsistency, where the minimum $\chi^2$ per degree of freedom of the fitting is larger than 4. In addition, the full NLO NRQCD calculation has difficulties to explain the polarization of exited state $\psip$ measured at Tevatron \cite{Abulencia:2007us}, as well as the polarization of heavier quarkonium, such as $\Upsilon(3S)$ measured by CMS at the LHC \cite{Chatrchyan:2012woa,Gong:2013qka}, although it is capable of explaining the data on the $\jpsi$ polarization \cite{Chao:2012iv}.
Because of the large logarithms and possible huge power enhancement at higher orders, it is difficult to determine whether such inconsistencies are from large high order corrections or from the failure of NRQCD factorization conjecture.

Recently, a new QCD factorization approach to high $p_T$ heavy quarkonium production at collider energies was proposed \cite{Kang:2011mg,Kang:2011zza,KMQS-hq1,KMQS-hq2}.  A similar factorization approach based on soft-collinear effective theory (SCET) was also proposed \cite{Fleming:2012wy}. In the QCD factorization approach, the cross section is first expanded by powers of $1/p_T^2$. As argued in Refs.~\cite{KMQS-hq1,KMQS-hq2}, both the leading-power (LP) term and next-to-leading-power (NLP) term of the expansion could be factorized systematically into infrared-safe short-distance partonic hard parts convoluted with universal fragmentation functions (FFs), plus parton distribution functions (PDFs) in the case of hadronic collisions.  Unlike the NRQCD factorization approach, the short-distance hard parts calculated by using the QCD factorization formalism are free of large logarithms and the power enhancements.  All powers of $\mylog{p_T^2/m_Q^2}$-type logarithms are resumed by solving a closed set of evolution equations of FFs.  Because of its better control on high order corrections, the QCD factorization approach is a powerful tool to check our understanding of heavy quarkonium production.

Similar to the inclusive production of a light hadron at high $p_T$, the predictive power of QCD factorization approach to heavy quarkonium production requires our knowledge of the FFs, in addition to the systematically calculated short-distance partonic hard parts.  With the perturbatively calculated evolution kernels, we only need the FFs at an input scale $\mu_0 \gtrsim 2m_Q$, while the evolution equations could generate the FFs to any other scales.  However, because of the inclusion of NLP contribution to the cross section, it requires more unknown input FFs.    For the LP term, we need a minimum of {\it two} single parton (light quark + gluon) FFs to each heavy quarkonium state, if we assume that all light quark/antiquark flavors share the same FF, plus one or two more input FFs if we include fragmentation contribution from a heavy quark, whose mass $m_Q\ll p_T$.  For the NLP term, we need at least {\it six} heavy quark-pair FFs due to the pair's two color and four spin states (vector, axial vector, and tensor states), if we do not distinguish the two tensor states.  Combining the LP and NLP contributions, we need a minimum of {\it eight} to {\it ten}  unknown input FFs to describe the production of each heavy quarkonium state. Although contributions from some fragmentation channels, such as the tensor channels, could be less important, it still requires a lot of information/data to extract these FFs, which makes it difficult to test this factorization formalism precisely.

Like all QCD factorization approaches to high $p_T$ hadron production, it is the FFs at the input scale that are most sensitive to the properties of the heavy quarkonium produced, since the perturbatively calculated partonic hard parts and evolution kernels are insensitive to any long-distance characteristics, such as the spin and polarization, of the produced quarkonium.  That is, the knowledge of heavy quarkonium FFs at the input scale is extremely important for understanding the production and formation of different heavy quarkonia at collider energies.

Unlike the light hadron FFs, heavy quarkonium FFs have an intrinsic hard scale - the heavy quark mass $m_Q$, which could be large enough to be considered as a perturbative scale. With the input scale $\mu_0\gtrsim 2m_Q$, NRQCD could be a good effective theory to be used to calculate these unknown input FFs, because they do not have large logarithmic terms or the huge power enhancement at $\mu_0$.
Although there is no formal proof that NRQCD factorization works for evaluating these universal input FFs perturbatively to all orders in $\alpha_s$ and all powers in expansion of heavy quark velocity, $v$, it was demonstrated \cite{Nayak:2005rw,Nayak:2006fm} that NRQCD factorization works up to two-loop radiative corrections.  It was proposed to use NRQCD factorization to evaluate heavy quarkonium FFs at the input scale \cite{Kang:2011mg,KMQS-hq2}, as a conjecture, so that all unknown heavy quarkonium input FFs could be given by functions of a few unknown, but, universal NRQCD LDMEs with the perturbatively calculated short-distance coefficients in terms of NRQCD factorization.

In this paper and a companion paper \cite{\Pwave}, we present our calculation of the input heavy quarkonium FFs at the scale $\mu_0$ using NRQCD factorization approach, including contributions through both $S$-wave and $P$-wave NRQCD $\cc$-states. We derive the FFs from a perturbatively produced heavy quark pair for all partonic channels at ${\cal O}(\alpha_s^0)$ and ${\cal O}(\alpha_s)$.  For the completeness, we also present the FFs from a single parton at both ${\cal O}(\alpha_s)$ and ${\cal O}(\alpha_s^2)$.  All heavy quarkonium FFs from our calculation have an explicit and definite dependence on momentum fractions and the input factorization scale $\mu_0$, which should be a parameter to be determined by fitting experimental data, along with a few unknown NRQCD LDMEs for each physical heavy quarkonium state.  From the existing phenomenological success of NRQCD factorization approach to inclusive production of heavy quarkonia, and the clear separation of momentum scales, we expect that our results should provide a reasonable description of these non-perturbative FFs at the input scale.  With our calculated input FFs, the evolution kernels of FFs in Ref.~\cite{KMQS-hq1}, and the short-distance perturbative hard parts from Ref.~\cite{KMQS-hq2}, we should be able to perform the first numerical predictions for heavy quarkonium production at collider energies in terms of the QCD factorization approach, which is beyond the scope of this paper.

Within the frame work of NRQCD factorization approach, these input heavy quarkonium FFs could be systematically improved in powers of both coupling constant $\alpha_s$ and heavy quark velocity $v$.  We are aware that without a formal proof of NRQCD factorization for calculating these FFs, some modifications to these nonperturbative FFs might be needed for a better description of data.  In this sense, any calculation in QCD factorization approach to heavy quarkonium production by using our calculated input FFs is a good test of NRQCD factorization. Any modification to our calculated input FFs for a better description of data may provide insight to the validity of NRQCD factorization.

The rest of this paper is organized as follows. In section \ref{sec:pQCDFac}, we briefly review the QCD factorization approach to heavy quarkonium production at collider energies. In section \ref{sec:ApplyNRQCD}, we apply NRQCD factorization to heavy quarkonium FFs from a single parton as well as a heavy quark pair. Since various single parton to heavy quarkonium fragmentation functions are available in the literature, we concentrate on the detailed calculations of FFs from a heavy quark pair, while we provide our full results of the FFs from a single parton in Appendix \ref{app:SinglePartonFF} with a brief discussion.  We present our leading order (LO) and NLO calculation of the FFs from a heavy quark-pair by using an example $[\cc(\aeight)] \to [\cc(\COaSz)]$ in Section \ref{sec:LO} and \ref{sec:NLO}, respectively. Our complete results for heavy quarkonium FFs through a $S$-wave NRQCD $\cc$-states are listed in Appendix \ref{app:doubleresults}.  Our conclusions are summarized in Section \ref{sec:summary}.  Calculation details and full results for the FFs through a $P$-wave NRQCD $\cc$-states are presented in a companion paper \cite{\Pwave}.


\section{QCD factorization approach}
\label{sec:pQCDFac}

In the QCD factorization approach, the production cross section of a heavy quarkonium $H$ with momentum $p$ at a large transverse momentum $p_T$ in the lab frame is expanded in a power series of $1/p_T$ \cite{Kang:2011mg,Kang:2011zza,KMQS-hq1,KMQS-hq2}
\begin{linenomath*}
\begin{align}\label{eq:pQCDfac}
\begin{split}
E_p \frac{d\sigma_{A+B\to H+X}}{d^3p}(p)
&\approx
\sum_f\int\frac{dz}{z^2}D_{f\to H}(z;m_Q)
E_c\frac{d\hat{\sigma}_{A+B\to f(p_c)+X}}{d^3p_c}\Big(p_c=\frac{1}{z}\hat{p}\Big)
\\
&\hspace{0cm}
+\sum_{[Q\bar{Q}(\kappa)]}\int\frac{dz}{z^2}\, \frac{d\zeta_1\, d\zeta_2}{4}\, {\cal{D}}_{[Q\bar{Q}(\kappa)]\to H}(z,\zeta_1,\zeta_2;m_Q)
\\
&\hspace{1.5cm}
 \times
 E_c\frac{d\hat{\sigma}_{A+B\to[Q\bar{Q}(\kappa)](p_c)+X}}{d^3 p_c}
(P_Q,
P_{\bar{Q}};
P'_Q,
P'_{\bar{Q}}),
\end{split}
\end{align}
\end{linenomath*}
where the factorization scale $\mu_F$ dependence is suppressed, and
the summation over unobserved particles $X$ is understood.
In Eq.~(\ref{eq:pQCDfac}), the heavy quarkonium momentum
$p^\mu$ is defined in the lab frame as $p^\mu = (m_T \cosh y,\, {\bf{p}}_T,\, m_T \sinh y)$
with rapidity $y$, $m_T=\sqrt{m_H^2 +p_T^2}$ and $p_T=\sqrt{{\bf p}_T^2}$.
For our calculation of input FFs, it is more convenient to define the momentum $p^\mu$
in a frame in which it has no transverse component as $p^\mu = (p^+,\, p^-\,, 0_\perp)$ with
\begin{eqnarray}
p^+ &=& \left(m_T\,\cosh y+\sqrt{p_T^2+m_T^2\,\sinh^2 y}\right)/\sqrt{2}\, ,
\nonumber\\
p^- &=& \left(m_T\,\cosh y-\sqrt{p_T^2+m_T^2\,\sinh^2 y}\right)/\sqrt{2}\, ,
\label{eq:lighcone:p}
\end{eqnarray}
in terms of the rapidity and transverse momentum in the lab frame.
The components in the light-cone coordinate in Eq.~(\ref{eq:lighcone:p}) are defined as
$p^\pm = (p^0\pm p^3)/\sqrt{2}$.
With the two light-like vectors $\hat{\bar{n}}^\mu = (1^+, 0^-, 0_\perp)$
and $\hat{n}^\mu = (0^+, 1^-, 0_\perp)$, which satisfy
$\hat{\bar{n}}^2=\hat{n}^2=0$ and $\hat{\bar{n}}\cdot \hat{n} = 1$, the light-cone components
of momentum $p^\mu$ can be expressed as $p^+ = p\cdot \hat{n}$ and $p^- = p \cdot \hat{\bar{n}}$.
In this frame, we have the momenta of perturbatively produced
partons in Eq.~(\ref{eq:pQCDfac}) as
$p_c = \hat{p}/z$ with $\hat{p}^\mu = (p^+,\, 0^-, 0_\perp)=p^\mu(m_H=0)$
(or $z=\hat{p}^+/p_c^+$), and
\begin{linenomath*}
\begin{align}\label{eq:mom}
\begin{split}
P_Q=\frac{1+\zeta_1}{2}\,p_c,\,\,
P_{\bar{Q}} = \frac{1-\zeta_1}{2}\,p_c,\,\,
P'_Q= \frac{1+\zeta_2}{2}\,p_c,\,\,
P'_{\bar{Q}}=\frac{1-\zeta_2}{2}\,p_c\, ,
\end{split}
\end{align}
\end{linenomath*}
where $\zeta_1$ and $\zeta_2$ are relative light-cone momentum fractions between the heavy quark and antiquark in the amplitude and its complex conjugate, respectively.
Note that in Eq.~(\ref{eq:pQCDfac}), we used variables $\zeta_1$ and $\zeta_2$ instead of
the $u$ and $v$ used in Ref.~\cite{KMQS-hq1}, which are one-to-one corresponded as
$\zeta_1=2u-1$, $\zeta_2=2v-1$, and $d\zeta_1 d\zeta_2 /4 = du\, dv$.

\begin{figure}[htb]
\begin{center}
\hspace*{-0cm}
\includegraphics[width=0.45\textwidth,height=5.5cm]{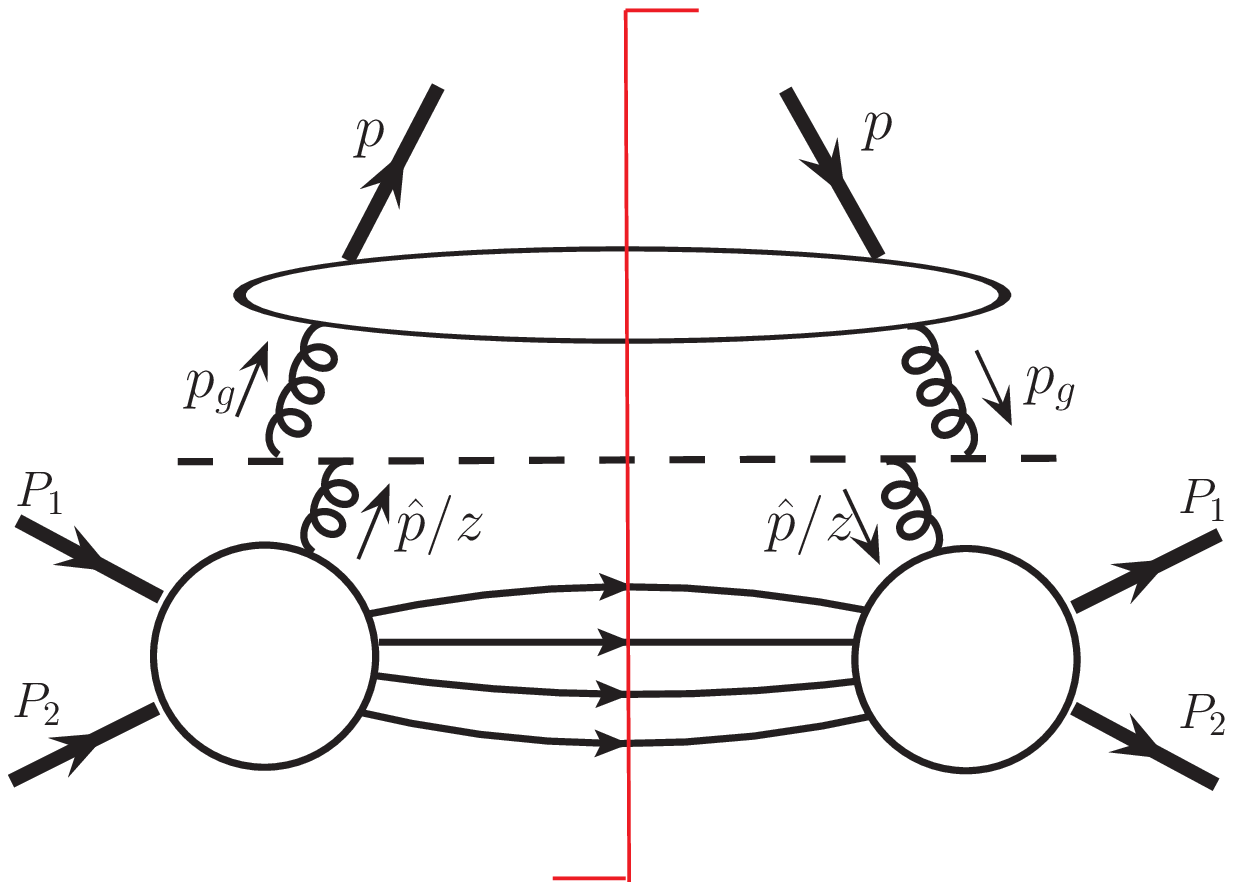}
\includegraphics[width=0.45\textwidth,height=5.5cm]{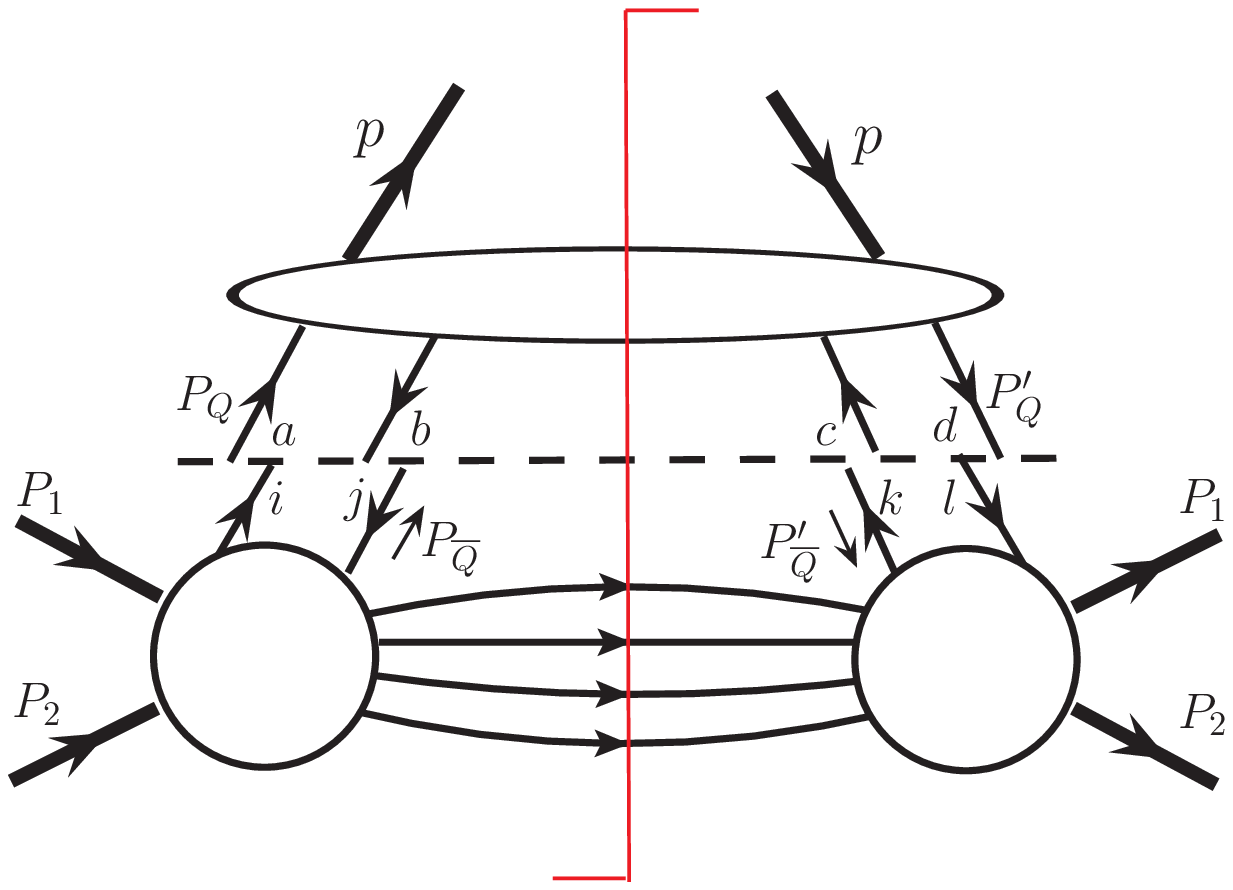}
\caption{pQCD factorization diagrams of heavy quarkonium production. Left: single parton (here taking gluon as an example) fragmentation; right: heavy quark pair fragmentation.
\label{fig:pQCDFac}}
\end{center}
\end{figure}

The factorization formula in Eq.~(\ref{eq:pQCDfac}) was argued to be valid in QCD perturbation theory to all orders in $\as$ \cite{KMQS-hq1}.  The first term on the right-hand-side is the leading power (LP) contribution to the production cross section in its $1/p_T$ expansion, while the second term is the next-to-leading power (NLP) contribution, or the first power correction.  The Feynman diagrams in the cut diagram notation for these two terms are shown in Fig. \ref{fig:pQCDFac}.
Physically, the first term represents the production of a single parton of flavor $f$ at short-distance, followed by its fragmentation into the observed heavy quarkonium $H$.
The $\sum_f$ runs over all parton flavors $f=q,\bar{q}, g$ including heavy quarks with its mass $m_Q\ll p_T$. For collider energies at the LHC, the sum could include charm quark $c$ as well as bottom quark $b$.
The second term describes the production of a heavy $\cc$-pair at the hard collision, and the pair then
fragments into an observed heavy quarkonium $H$.
The $\sum_{[Q\bar{Q}(\kappa)]}$ runs over all possible spin and color states of the $\cc$-pair, which could be the vector ($v^{[1, 8]}$), axial-vector ($a^{[1, 8]}$) or tensor ($t^{[1, 8]}$) state, with the superscripts labeling the color state of the pair: singlet ($^{[1]}$) or octet ($^{[8]}$).
The projection operators of different $\cc$-pair states are given in Ref.~\cite{KMQS-hq1}.
For completeness, we also list these operators in Appendix \ref{app:proj}.
Note that in the diagram on the right in Fig. \ref{fig:pQCDFac}, the $\cc$-pair on the left of the cut could have different relative momentum from the $\cc$-pair on the right, which means that $\zeta_1$ is not necessarily equal to $\zeta_2$ in Eq.~(\ref{eq:pQCDfac}).

In Eq.~(\ref{eq:pQCDfac}), the short-distance partonic hard parts $d\hat{\sigma}$ could be systematically calculated in powers of $\as$ (needs to convolute with parton distribution functions (PDFs)
if $A$ and/or $B$ is a hadron).
The fragmentation functions $D_{f\to H}(z;m_Q,\mu_F)$ and ${\cal{D}}_{[Q\bar{Q}(\kappa)]\to H}(z,\zeta_1,\zeta_2;m_Q,\mu_F)$ are unknown, but process independent, universal functions.
Their dependence on factorization scale $\mu_F$ is determined by a closed set of evolution equations \cite{KMQS-hq1},
\begin{linenomath*}
\begin{align}
\begin{split}\label{eq:evo_1}
&\frac{\partial}{\partial\ln\mu_F^2} D_{f\to H}(z;m_Q,\mu_F)
=
\sum_{f'}
\int_{z}^1 \frac{dz'}{z'}
{D}_{f'\to H}(z';m_Q, \mu_F)\ \gamma_{f\to f'} (z/z',\alpha_s)
\\
&\hspace{3cm}
+
\frac{1}{\mu_F^2}\
\sum_{[Q\bar{Q}(\kappa')]}
\int_{z}^1 \frac{dz'}{z'} \int_{-1}^1 \frac{d\zeta'_1}{2}  \int_{-1}^1 \frac{d\zeta'_2}{2} \,
{\mathcal D}_{[Q\bar{Q}(\kappa')]\to H}(z', \zeta'_1,  \zeta'_2; m_Q, \mu_F)
\\
&\hspace{6cm}
\times
\gamma_{f\to [Q\bar{Q}(\kappa')]}(\frac{z}{z'}, u'=\frac{1+\zeta'_1}{2}, v'=\frac{1+\zeta'_2}{2}) ,
\end{split}\\
\begin{split}\label{eq:evo_2p}
&
\frac{\partial}{\partial{\ln\mu_F^2}}{\mathcal D}_{[Q\bar{Q}(\kappa)]\to H}(z, \zeta_1,\zeta_2;m_Q, \mu_F)
\\
&\hspace{1.5cm}
=
\sum_{[Q\bar{Q}(\kappa')]}
\int_{z}^1 \frac{dz'}{z'}
\int_{-1}^1 \frac{d\zeta'_1}{2}  \int_{-1}^1 \frac{d\zeta'_2}{2}\,
{\mathcal D}_{[Q\bar{Q}(\kappa')]\to H}(z', \zeta'_1, \zeta'_2;m_Q, \mu_F)
\\
&\hspace{2cm}
\times
\Gamma_{[Q\bar{Q}(\kappa)]\to [Q\bar{Q}(\kappa')]}(\frac{z}{z'}, u=\frac{1+\zeta_1}{2}, v=\frac{1+\zeta_2}{2}; u'=\frac{1+\zeta'_1}{2}, v'=\frac{1+\zeta'_2}{2}) ,
\end{split}
\end{align}
\end{linenomath*}
where we explicitly convert the variables $u$ and $v$ to $\zeta_1$ and $\zeta_2$ in the argument of evolution kernels $\gamma_{f\to [Q\bar{Q}(\kappa')]}$ and $\Gamma_{[Q\bar{Q}(\kappa)]\to [Q\bar{Q}(\kappa')]}$ to avoid confusion.  The evolution kernels $\gamma's$ and $\Gamma's$ are process-independent and perturbatively calculable. The well-known DGLAP evolution kernels $\gamma_{f\to f'}$ are available to next-to-next-to-leading order in $\as$. The power-mixing evolution kernels $\gamma_{f\to [Q\bar{Q}(\kappa')]}$
were calculated in Ref.~\cite{KMQS-hq1}, and the heavy quark pair evolution kernels
$\Gamma_{[Q\bar{Q}(\kappa)]\to [Q\bar{Q}(\kappa')]}$ have been recently calculated by two groups independently~\cite{Fleming:2013qu,KMQS-hq1}. If both $\kappa$ and $\kappa'$ are color singlet, the kernel $\Gamma_{[Q\bar{Q}(\kappa)]\to [Q\bar{Q}(\kappa')]}$ reduces to the well-known Efremov-Radyushkin-Brodsky-Lepage evolution kernel for exclusive processes~\cite{Lepage:1979zb,Efremov:1978rn}.

Similar to the FFs for pion or kaon production, a set of single parton and $\cc$-pair fragmentation functions at an input factorization scale $\mu_0$ is required as the boundary conditions (BCs) for solving the evolution equations in Eqs.~(\ref{eq:evo_1}) and (\ref{eq:evo_2p}).  For production of each heavy quarkonium state at high $p_T \gg m_Q$, we need {\it four} single parton input FFs and {\it six} $\cc$-pair input FFs as the required BCs.  Since these BCs are nonperturbative, in principle, they should be extracted from data.
However, extracting {\it ten} or more unknown functions for each physical heavy quarkonium is difficult in practice.  The extraction is practically feasible if we have some knowledge of these BCs, such as their functional forms.

When the factorization scale $\mu_F\to \mu_0\gtrsim 2m_Q$,
$\ln(\mu_0^2/m_Q^2)$-type logarithms as well as powers of $\mu_0^2/m_Q^2$
in NRQCD calculations are no longer large.
With a clear separation of momentum scales, $\mu_0 \sim O(m_Q) \gg m_Q v$,
NRQCD might be the right effective theory for calculating these input FFs
by factorizing the dynamics at $\mu_0$ from non-perturbative soft physics at the scale
of $m_Q v$ and below.
In the rest of this paper, as a conjecture \cite{Kang:2011mg,Kang:2011zza,KMQS-hq2},
we apply NRQCD factorization to these input FFs at $\mu_0$, and
calculate corresponding short-distance coefficient functions to the first nontrivial order in $\alpha_s$
for the fragmentation via all S-wave NRQCD $\cc$ states.



\section{NRQCD factorization for FF\lowercase{s}}
\label{sec:ApplyNRQCD}

In this section, we set up the prescription for calculating the heavy quarkonium FFs
at an input factorization scale, $\mu_0$, in terms of NRQCD factorization formalism \cite{Bodwin:1994jh}.

\subsection{Calculation of single-parton FFs}

We can write the NRQCD factorization formalism for
heavy quarkonium FFs from single parton as
\cite{Kang:2011mg,Kang:2011zza,KMQS-hq2}
\begin{linenomath*}
\begin{align}\label{eq:singleFF}
\begin{split}
D_{f\to H}(z;m_Q,\mu_0)
=
\sum_{[\cc(n)]} \hat{d}_{f\to[\cc(n)]}(z;m_Q,\mu_0,\mu_\Lambda)
\langle \mathcal{O}_{[\cc(n)]}^{H}(\mu_\Lambda)\rangle,
\end{split}
\end{align}
\end{linenomath*}
where $H$ represents a particular physical heavy quarkonium state,
$\mu_0\gtrsim 2m_Q$ represents the input QCD factorization scale at which
the $\ln(\mu_0/m_Q)$-type logarithmic contributions to the production cross section
are comparable with the $m_Q/\mu_0$-type power suppressed contribution,
$\mu_\Lambda \sim m_Q$ is NRQCD factorization scale and does not have to be equal to $\mu_0$.
The summation runs over all intermediate non-relativistic $\cc$ states,
which is labelled as $n=\state{{2S+1}}{L}{J}{1,8}$, with superscript $^{[1]}$ (or $^{[8]}$) denoting color singlet (or octet) state.
Short-distance coefficients $\hat{d}_{f\to[\cc(n)]}(z;m_Q,\mu_0,\mu_\Lambda)$
describe the dynamics at energy scale larger than $\mu_\Lambda \gg \Lambda_{QCD}$,
thus they could be calculated perturbatively.
LDMEs $\langle \mathcal{O}_{[\cc(n)]}^{H}(\mu_\Lambda)\rangle$
include all interactions below scale $\mu_\Lambda$, and are intrinsically nonperturbative.
These universal LDMEs are scaled in powers of $\cc$-pair's relative velocity $v\ll 1$ in the rest frame of $H$.
Hence, in practice, the summation could be approximately truncated, with only a few terms left to be considered.
For example, to calculate $J/\psi$ production at the LHC, the most important LDMEs are $n=\CScSa,\COaSz, \COcSa$ and $\COcPj$ up to order $v^4$.  In Eq.~(\ref{eq:singleFF}),
the factorization scales, $\mu_0$ and $\mu_{\Lambda}$,
along with the LDMEs, should be determined by fitting experimental data.

Since the short-distance coefficients $\hat{d}_{f\to[\cc(n)]}(z;m_Q,\mu_0,\mu_\Lambda)$ are not sensitive to long-distance details of the heavy quarkonium state, the same factorization formula in Eq.~(\ref{eq:singleFF}) could be applied to an asymptotic partonic state, such as an asymptotic $\cc$-pair state. By replacing the heavy quarkonium state $H$ with an asymptotic $\cc$-pair state, $[\cc(n')]$, we can write
\begin{linenomath*}
\begin{align}\label{eq:singleFF0}
\begin{split}
D_{f\to [\cc(n')]}(z;m_Q,\mu_0)
=
\sum_{[\cc(n)]} \hat{d}_{f\to[\cc(n)]}(z;m_Q,\mu_0,\mu_\Lambda)
\langle \mathcal{O}_{[\cc(n)]}^{[\cc(n')]}(\mu_\Lambda)\rangle.
\end{split}
\end{align}
\end{linenomath*}
With this form, one could calculate the $D_{f\to [\cc(n')]}(z;m_Q,\mu_0)$ on the left with perturbative QCD and
the $\langle \mathcal{O}_{[\cc(n)]}^{[\cc(n')]}(\mu_\Lambda)\rangle$ on the right
with perturbative NRQCD.
If NRQCD factorization is valid for these input FFs,
the LDMEs on the right should reproduce all infrared (IR) and
Coulomb divergences in $D_{f\to [\cc(n')]}(z; m_Q,\mu_0)$, with short-distance coefficients
$\hat{d}_{f\to[\cc(n)]}(z;m_Q,\mu_0,\mu_\Lambda)$ IR-safe to all orders.

However, there is a major difference between applying NRQCD factorization to the heavy quarkonium production cross sections and to the heavy quarkonium FFs \cite{KMQS-hq2}.
For the production cross section, all perturbative UV divergences are completely taken care of by the renormalization of QCD.  For the input FFs, on the other hand, there are additional perturbative UV divergences associated with the composite operators that define the FFs.   Since NRQCD factorization in the right-hand-side (RHS) of Eq.~(\ref{eq:singleFF}), so as Eq.~(\ref{eq:singleFF0}), is a factorization of soft region corresponding to heavy quark binding, it does not deal with the UV divergence of the composite operators defining the FFs in the left-hand-side (LHS) of the same equation.  That is, the matching in Eq.~(\ref{eq:singleFF}), so as in Eq.~(\ref{eq:singleFF0}), and similarly, that in Eq.~(\ref{eq:QQFFNR1}) below, makes sense only if all perturbative UV divergences associated with the composite operators defining the FFs in the left-hand-side (LHS) are renormalized and any ambiguity in connection with this renormalization is simply a part of factorization scheme dependence of the FFs \cite{KMQS-hq2}.

Although a formal proof for the NRQCD factorization formula in Eq.~(\ref{eq:singleFF}) is still lacking,
The derivation of the coefficients $\hat{d}_{f\to[\cc(n)]}(z;m_Q,\mu_0,\mu_\Lambda)$ by
calculating both sides of Eq.~(\ref{eq:singleFF0}) perturbatively actually provides an explicit verification
of the factorization formalism, order-by-order in perturbation theory.
In the case of single parton FFs, we calculated all the short-distance coefficients up to ${\cal O}(\as^2)$ and no inconsistency has been found.
Many of these single parton FFs have been calculated before and are available in the literature \cite{Braaten:2000pc,Ma:1995vi,Braaten:1996rp,Braaten:1994kd,Braaten:1993mp,Hao:2009fa,Braaten:1993rw,Jia:2012qx}.
We found that our results agree with almost all of them.
Since enough calculation details were presented in those early papers,
here we simply list our complete results for single parton FFs in Appendix \ref{app:SinglePartonFF},
and point out any differences from early publications.


\subsection{Calculation of $\cc$-pair FFs}

The fragmentation function for a $\cc$-pair in a particular spinor and color state $\kappa$ to a physical heavy quarkonium $H$ with momentum $p$ is defined as \cite{KMQS-hq1}
\begin{linenomath*}
\begin{align}\label{eq:QQFF}
\begin{split}
&{\cal D}_{[\cc(\kappa)]\to H}
(z,\zeta_1,\zeta_2;m_Q, \mu_0 )
=\int\frac{p^+ dy^-}{2\pi}\frac{p^+/z \,dy_1^-}{2\pi}\frac{p^+/z \,dy_2^-}{2\pi}
\\
&\hspace{3cm}
\times
e^{-i(p^+/z)y^-}e^{i(p^+/z)[(1-\zeta_2)/2]\,y_1^-}e^{-i(p^+/z)[(1-\zeta_1)/2]\,y_2^-}\\
&\hspace{3cm}
\times {\cal P}_{ij,kl}^{(s)}(p_c)\, {\cal C}_{ab,cd}^{[I]}\,
\langle 0 |\bar{\psi}_{c',k}(y_1^-) [\Phi_{\hat{n}}^{(F)}(y_1^-)]_{c'c}^\dag [\Phi_{\hat{n}}^{(F)}(0)]_{d d'}\, \psi_{d',l}(0) |H(p)\, X\rangle\\
&\hspace{3cm}
\times
\langle H(p)\,X|
\bar{\psi}_{a',i}(y^-) [\Phi_{\hat{n}}^{(F)}(y^-)]^\dag_{a' a} [\Phi_{\hat{n}}^{(F)}(y^-+y_2^-)]_{b b'} \psi_{b',j}(y^-+y_2^-) |0\rangle,
\end{split}
\end{align}
\end{linenomath*}
where subscripts $i,j,k,l$ are the spinor indices of heavy (anti-)quark fields,
$a,a',b,b'\ldots$ are color indices of SU(3) color in the fundamental representation,
and the summation over repeated indices are understood.
Operators ${\cal P}_{ij,kl}^{(s)}(p_c)$ and ${\cal C}_{ab,cd}^{[I]}$ project the fragmenting $\cc$-pair
to a particular spin and color state $\kappa$,
which could be a vector ($v^{[1,8]}$), axial-vector ($a^{[1,8]}$)
or tensor ($t^{[1,8]}$) state, with superscript denoting the color.
Definitions of these projection operators are listed in Appendix \ref{app:proj}.
Since the relative momenta of the $\cc$-pairs in the amplitude and its complex conjugate are not necessarily the same, $\zeta_1$ and $\zeta_2$ could be different.
$\Phi_{\hat{n}}^{(F)}$ is the gauge link to make the fragmentation function gauge invariant, and is defined as
\begin{linenomath*}
\begin{align}\label{eq:GLdef}
\begin{split}
\Phi_{\hat{n}}^{(F)}(y^-)={\cal P}\,\text{exp}\left[-i\,g \int_{y^-}^{\infty} d\lambda \,{\hat{n}}\cdot A^{(F)}(\lambda {\hat{n}})\right],
\end{split}
\end{align}
\end{linenomath*}
where superscript $(F)$ indicates the fundamental representation.

Assuming that NRQCD factorization works for heavy quarkonium FFs, we can factorize the heavy quarkonium FFs from a $\cc$-pair as \cite{KMQS-hq2}
\begin{linenomath*}
\begin{align}\label{eq:QQFFNR1}
\begin{split}
{\cal D}_{[\cc(\kappa)]\to H}(z,\zeta_1,\zeta_2; m_Q, \mu_0)
=
\hspace{-0.15cm}
\sum_{[\cc(n)]}
\hspace{-0.15cm}
 \hat{d}_{[\cc(\kappa)]\to[\cc(n)]}(z,\zeta_1,\zeta_2; m_Q,\mu_0,\mu_\Lambda)
\langle \mathcal{O}_{[\cc(n)]}^{H}(\mu_\Lambda)\rangle,
\end{split}
\end{align}
\end{linenomath*}
where symbols have the same meaning as those in Eq.~(\ref{eq:singleFF}).
If the factorization formalism in Eq.~(\ref{eq:QQFFNR1}) is valid, it should also be valid if we replace the heavy quarkonium state $H$ by any asymptotic partonic state.  By replacing the heavy quarkonium state $H$ with an asymptotic $\cc$-pair state, $[\cc(n')]$, we can write
\begin{linenomath*}
\begin{align}\label{eq:QQFFNR2}
\begin{split}
{\cal D}_{[\cc(\kappa)]\to [\cc(n')]}(z,\zeta_1,\zeta_2;m_Q,\mu_0)
=
\sum_{[\cc(n)]} \hat{d}_{[\cc(\kappa)] \to[\cc(n)]}(z,\zeta_1,\zeta_2;m_Q,\mu_0,\mu_\Lambda)
\langle \mathcal{O}_{[\cc(n)]}^{[\cc(n')]}(\mu_\Lambda)\rangle\, ,
\end{split}
\end{align}
\end{linenomath*}
and derive the short-distance coefficients, $\hat{d}_{[\cc(\kappa)]\to[\cc(n)]}(z,\zeta_1,\zeta_2;m_Q,\mu_0,\mu_\Lambda)$ above by calculating both sides of the equation, perturbatively.  If the factorization is valid,
any IR sensitivity of the fragmentation function to an asymptotic state of a $\cc$-pair on the left of the equation should be systematically absorbed into the NRQCD LDMEs on the right, in the same manner as in Eq.~(\ref{eq:singleFF0}).  As explained in the last subsection, the matching in Eq.~(\ref{eq:QQFFNR1}),
so as Eq.~(\ref{eq:QQFFNR2}), is possible only if the UV renormalization of the composite operators defining the FFs in the LHS of the equation is taken care of \cite{KMQS-hq2}.

In this paper, we use dimensional regularization to regularize various divergences involved in our NLO calculations.  With the definition in Eq.~(\ref{eq:QQFF}), we have an explicit $D$-dimensional expression
for the LHS of Eq.~(\ref{eq:QQFFNR2}) as
\begin{linenomath*}
\begin{align}\label{eq:QQFFQQ}
\begin{split}
\hspace{1.5cm}&\hspace{-1.5cm}
{\cal D}_{[\cc(s^{[b]})]\to [\cc(i^{[b']})]}(z,\zeta_1,\zeta_2;m_Q,\mu_0)
=\frac{z^{D-2}}{N_s N_b N_{i}^\text{NR} N_{b'}^\text{NR}}
\int\frac{{\text d}^{D} p_c}{(2\pi)^{D}}
\left(\prod_X\int\frac{{\text d}^{D-1} p_X}{(2\pi)^{D-1} 2E_X}\right)
\\
&
\times
(2\pi)^D\delta^D(p_c-p-\sum_X p_X)\delta(z-\frac{p^+}{p_c^+})\,
{\cal M}_{[\cc(s^{[b]})]\to [\cc(i^{[b']})]}(p, z, \zeta_1, \zeta_2)
+\mbox{UVCT}(\mu_0)
\\
=&
\frac{z^{D-2}}{N_s N_b  N_{i}^\text{NR} N_{b'}^\text{NR}}
\left(\prod_X\int\frac{{\text d}^{D-1} p_X}{(2\pi)^{D-1} 2E_X}\right)
\delta(z-\frac{p^+}{p_c^+})
\\
&\times
{\cal M}_{[\cc(s^{[b]})]\to [\cc(i^{[b']})]}(p, z, \zeta_1, \zeta_2)
+\mbox{UVCT}(\mu_0)\, ,
\end{split}
\end{align}
\end{linenomath*}
where $p$ is the momentum of produced heavy quark state $[\cc(i^{[b']})]$,
and ``UVCT$(\mu_0)$'' indicates the UV counter-term
needed to remove the UV divergence associated with the composite operators defining the FFs.
In Eq.~(\ref{eq:QQFFQQ}), we have separated the spinor and color labels for both initial and final $\cc$-pair.
$s$ and $b$ ($i$ and $b'$) denote the spin and color state for the incoming (outgoing) $\cc$-pair.
$s$ could be vector ($v$), axial-vector ($a$) or tensor ($t$).
$i$ is labelled with spectroscopic notation $^{2S+1}L_J$.
Color state $b$ and $b'$ can be either $``1"$ for color singlet or $``8"$ for color octet.
$N_s$ and $N_b$ ($N_{i}^\text{NR}$ and $N_{b'}^\text{NR}$) are the spin and color normalization factors for the incoming (outgoing) $\cc$-pair.
Their definitions are listed in Appendix \ref{app:proj}.
The phase space integration for unobserved particles $X$  is given explicitly.

The function ${\cal M}$ in Eq.~(\ref{eq:QQFFQQ}) is given by
\begin{linenomath*}
\begin{align}\label{eq:QQFFQQM2}
\begin{split}
{\cal M}_{[\cc(s^{[b]})]\to [\cc(i^{[b']})]}(p, z, \zeta_1, \zeta_2)
=&\,
{\text{Tr}}\left[\Gamma_s(p_c)\, C_b\, {\cal A}_{[\cc(s^{[b]})]\to [\cc(i^{[b']})]}(p, z, \zeta_1)\right] \\
&
\hspace{-2.5cm}\times
{\text{Tr}}\left[\Gamma_s^\dag(p_c)\, C_b^\dag\, {\cal A}^\dag_{[\cc(s^{[b]})]\to [\cc(i^{[b']})]}(p, z,\zeta_2)\right]
\times P_s(p_c)\, P_{i}^\text{NR}(p)
\, ,
\end{split}
\end{align}
\end{linenomath*}
where ``$\text{Tr}$'' is understood as the trace for both spinor and color.
In deriving Eq.~(\ref{eq:QQFFQQM2}), we explicitly write the spinor (color) projection operator
${\cal P}^{(s)}(p_c)$ (and ${\cal C}^{[I]}$) in Eq.~(\ref{eq:QQFF}) as a product of corresponding operator
in the scattering amplitude and that in its complex conjugate, such that
${\cal P}^{(s)}(p_c)\equiv {\Gamma_s(p_c) \Gamma_s^\dag(p_c) P_s}/{N_s}$
(and ${\cal C}^{[I]}\equiv C_b C_b^\dag/N_b$).
All of these projection operators and corresponding normalization factors
are listed in Appendix \ref{app:proj}.

The transition amplitude ${\cal A}$ in Eq.~(\ref{eq:QQFFQQM2}) is defined as,
\begin{linenomath*}
\begin{align}\label{eq:QQFFQQA}
\begin{split}
{\cal A}_{[\cc(s^{[b]})]\to [\cc(i^{[b']})]}(p,z,\zeta_1)
=&
\lim_{q_r\to0}\left(\prod_{j=0}^L
\frac{\text{d}}{\text{d}{q_r^{\alpha_j}}}\right)
\left\{
\int\frac{{\text d}^D q_1}{(2\pi)^D}
\times
2\,\delta(\zeta_1-\frac{2q_1^+}{p_c^+})\,\right.\\
&
\qquad\qquad\times
{\cal \bar{A}}_{[\cc(s^{[b]})]\to [\cc(i^{[b']})]}(q_1,q_r)\,
\Gamma_{i}^{\text{NR}}(p)\, C_{b'}^\text{NR}
\bigg\}
\, ,
\end{split}
\end{align}
\end{linenomath*}
where $\bar{\cal A}$ is the amputated amplitude, and the factor 2 in front of the delta function comes from the integration of $y_1^-$ in Eq.~(\ref{eq:QQFF}).
Spin projection operators $\Gamma_{i}^{\text{NR}}$ and color projection operators $C_{b'}^\text{NR}$ for outgoing $Q$ and $\bar{Q}$ are defined in Appendix \ref{app:proj}, which may have Lorentz indexes and color indexes, respectively.
In Eq.~(\ref{eq:QQFFQQA}), $q_1$ ($q_r$) is the momentum of the incoming (outgoing) heavy quark
relative to the incoming (outgoing) $\cc$-pair's center of mass.
The derivative operation, $\prod_{j=0}^L {\text{d}}/{\text{d}{q_r^{\alpha_j}}}$,  
with $\alpha_j$ the Lorentz index of momentum $q_r$, 
picks up the contribution to the $L^{\text{th}}$ orbital angular momentum state, 
with $L=1,2,3\dots$ corresponding to the orbital angular momentum state 
$S,\,P,\, D\dots$ of final $\cc$-pair, respectively. 
For the contribution to a $S$-wave $Q\bar{Q}$ state,  $\prod_{j=0}^{L=0} {\text{d}}/{\text{d}{q_r^{\alpha_j}}}=1$, and there is no need for the derivative operation on $q_r$.  For higher orbital momentum states, $L>0$, we expand the amplitude to the $L^{\text{th}}$-order in $q_r$.  Note that the limit and derivative operation over $q_r$ in Eq.~(\ref{eq:QQFFQQA}), $\lim_{q_r\to0}\left(\prod_{j=0}^{L} {\text{d}}/{\text{d}{q_r^{\alpha_j}}}\right)$,  are outside of the $q_1$ integration.


\section{LO matching coefficients}\label{sec:LO}

In this section and the next section, we take process $[\cc(\aeight)]\to[\cc(\COaSz)]$ as an example to present our detailed calculation of ${\cal D}_{[\cc(s^{[b]})]\to [\cc(i^{[b']})]}(z,\zeta_1,\zeta_2;m_Q,\mu_0)$ and the extraction of $\hat{d}_{[\cc(\kappa)] \to[\cc(n)]}(z,\zeta_1,\zeta_2;m_Q,\mu_0,\mu_\Lambda)$.

The heavy quark pair FFs to a heavy quarkonium are defined in terms of heavy quark field operators in QCD, see Eq.~(\ref{eq:QQFF}) for example, while the heavy quark states in NRQCD factorization are defined as non-relativistic.  Therefore, there are matching coefficients between a fragmenting QCD heavy quark pair and a NRQCD heavy quark pair, defining the LDMEs.  We derive the LO matching coefficients for all heavy quark fragmentation channels in this section.

\begin{figure}[htb]
\begin{center}
\includegraphics[width=0.5\textwidth]{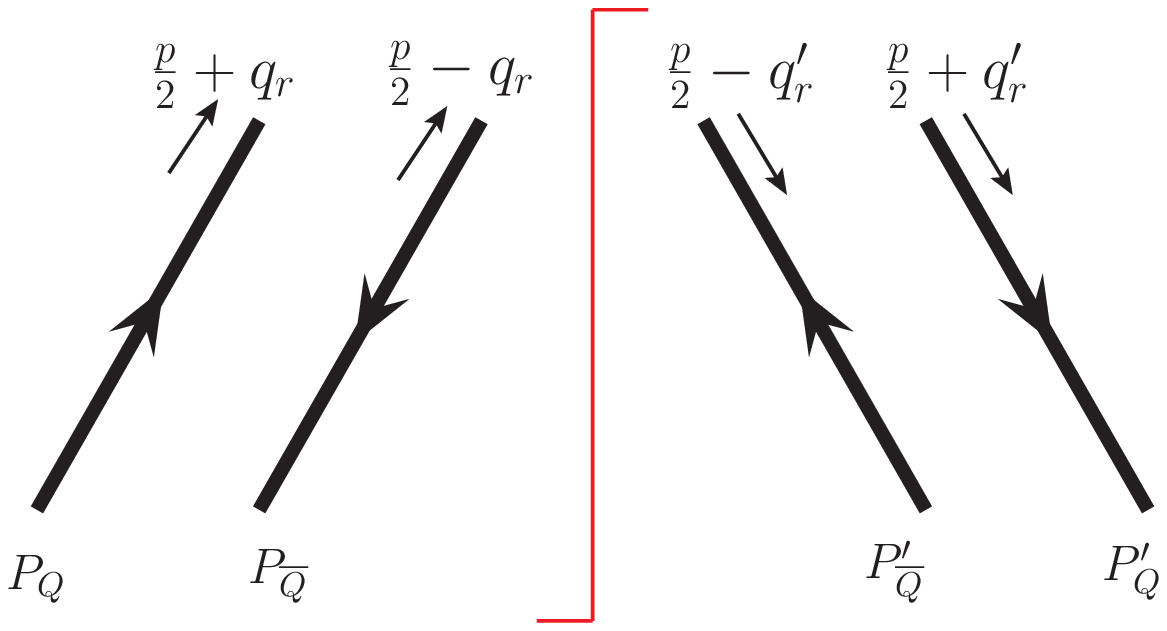}
\caption{Cut-diagram representation of ${\cal D}_{[\cc(s^{[b]})]\to [\cc(i^{[b']})]}(z,\zeta_1,\zeta_2;m_Q,\mu_0)$ at zeroth order.
\label{fig:LO}}
\end{center}
\end{figure}
A general cut-diagram representation for ${\cal D}_{[\cc(s^{[b]})]\to [\cc(i^{[b']})]}(z,\zeta_1,\zeta_2;m_Q,\mu_0)$ at zeroth order in power of $\as$ is given in Fig. \ref{fig:LO}, where momenta of incoming heavy quark and heavy antiquark are
defined as
\begin{linenomath*}
\begin{align}\label{eq:IQQmom}
\begin{split}
P_Q=\frac{p_c}{2}+q_1,\qquad P_{\bar{Q}}=\frac{p_c}{2}-q_1,\\
P'_Q=\frac{p_c}{2}+q_2,\qquad P'_{\bar{Q}}=\frac{p_c}{2}-q_2.
\end{split}
\end{align}
\end{linenomath*}
%
At the zeroth order, the LDME in Eq.~(\ref{eq:QQFFNR2}) is proportional to delta function $\delta_{n,n'}$. Thus, Eq.~(\ref{eq:QQFFNR2}) is simplified to
\begin{linenomath*}
\begin{align}\label{eq:QQFFNRmatchLO}
{\cal D}^{\text{LO}}_{[\cc(s^{[b]})]\to [\cc(i^{[b']})]}(z,\zeta_1,\zeta_2;m_Q,\mu_0)
= \hat{d}^{\text{ LO}}_{[\cc(s^{[b]})]\to [\cc(i^{[b']})]}(z,\zeta_1,\zeta_2;m_Q,\mu_0)\, .
\end{align}
\end{linenomath*}
Eqs.~(\ref{eq:QQFFQQ}) and (\ref{eq:QQFFQQA}) are reduced, respectively, to
\begin{linenomath*}
\begin{align}
&{\cal D}^{\text{LO}}_{[\cc(s^{[b]})]\to [\cc(i^{[b']})]}(z,\zeta_1,\zeta_2;m_Q,\mu_0)
=
\frac{\delta(1-z)}{N_s N_b N_{i}^\text{NR} N_{b'}^\text{NR}}
\,{\cal M}^{\text{LO}}_{[\cc(s^{[b]})]\to [\cc(i^{[b']})]}(p, z, \zeta_1, \zeta_2)\, ,\label{eq:QQFFQQLO}\\
&{\cal A}^{\text{LO}}_{[\cc(s^{[b]})]\to [\cc(i^{[b']})]}(p,z,\zeta_1)
=
\lim_{q_r\to0}
\left\{
2\,\delta(\zeta_1-\frac{2q_r^+}{p_c^+})
\right.\nonumber
\\
&
\hspace{5cm}
\times
\left.
{\cal \bar{A}}^{\text{LO}}_{[\cc(s^{[b]})]\to [\cc(i^{[b']})]}(q_1=q_r)
\Gamma_{i}^{\text{NR}}(p)\, C_{b'}^\text{NR}
\right\}.\label{eq:QQFFQQALO}
\end{align}
\end{linenomath*}
In Eq.~(\ref{eq:QQFFQQLO}), the delta function is expected because all momenta flow from incoming $\cc$-pair into the final $\cc$-pair.

One could further simplify the calculation by noting that at LO, the initial and final heavy quark pair must have {the same quantum numbers}, i.e. (1) color label $b$ and $b'$ must be the same; 
(2) spinor label $s$ and $i$ must have the same parity.
The parity of the outgoing $\cc$ state, $i=\statenc{{2S+1}}{L}{J}$ with $L=0$, is $(-1)^{S}$, 
while  the parity for the incoming $\cc$ state is $-1$ for $s=v,t$, and $+1$ for $s=a$.
Processes violating either of these two rules, such as ${\cal D}_{[\cc(s^{[1]})]\to [\cc(i^{[8]})]}(z,\zeta_1,\zeta_2;m_Q,\mu_0)$ and ${\cal D}_{[\cc(v^{[b]})]\to [\cc(\state{1}{S}{0}{b'})]}(z,\zeta_1,\zeta_2;m_Q,\mu_0)$, must vanish at this order.

For our example $[\cc(\aeight)]\to[\cc(\COaSz)]$, there is no derivative of $q_r$ in Eq.~(\ref{eq:QQFFQQALO}). From Eqs.~(\ref{eq:QQFFQQM2}) and (\ref{eq:QQFFQQALO}), we have
\begin{linenomath*}
\begin{align}\label{eq:QQFFQQLOSTr}
\begin{split}
&{\text{Tr}}\left[\Gamma_a(p_c)\, C_8\, {\cal A}^{\text{LO}}_{[\cc(\aeight)]\to [\cc(\COaSz)]}(p, z, \zeta_1)\right]
\\
=&
\text{Tr}_c\left[\sqrt{2}t^{(F)}_{c_{in}} \sqrt{2}t^{(F)}_{c_{out}}\right]
\text{Tr}_\gamma\left[\frac{\gamma\cdot \hat{n}\,\gamma_5 - \gamma_5\,\gamma\cdot \hat{n}}{8 p\cdot \hat{n}}\frac{1}{\sqrt{8m_Q^3}}(\frac{\slashed{p}}{2}-m_Q)\gamma_5(\frac{\slashed{p}}{2}+m_Q)\right]
\times 2\,\delta(\zeta_1)
\\
=&-\frac{1}{\sqrt{2m_Q}}\,\delta_{c_i,\, c_f}\,\delta(\zeta_1),
\end{split}
\end{align}
\end{linenomath*}
where ``$\text{Tr}_c$'' is the trace for color, ``$\text{Tr}_\gamma$" is the trace of $\gamma$-matrices, and $c_{in}$($c_{out}$) is the color for the incoming (outgoing) $\cc$-pair.  In Eq.~(\ref{eq:QQFFQQLOSTr}), we used the operator definitions given in Appendix \ref{app:proj} and the fact that $p_c=p$ for deriving the right-hand-side (RHS) of the equation.  For carrying out the trace of $\gamma$-matrices in Eq.~(\ref{eq:QQFFQQLOSTr}), we need to specify the definition of $\gamma_5$ in $D$-dimension.  Details of our prescription of $\gamma_5$ in $D$-dimension can be found in Appendix \ref{app:gamma5}.
The delta function $\delta(\zeta_1)$ indicates that the momenta of the initial heavy quark and heavy antiquark must be the same, since we have set the relative momentum of the final-state heavy quark and anti-quark to zero. Finally, combining the result Eq.~(\ref{eq:QQFFQQLOSTr}) with Eqs.~(\ref{eq:QQFFQQM2}), (\ref{eq:QQFFNRmatchLO}) and (\ref{eq:QQFFQQLO}), we obtain
\begin{linenomath*}
\begin{align}\label{eq:resLOS}
\begin{split}
\hat{d}^{\text{ LO}}_{[\cc(\aeight)]\to [\cc(\COaSz)]}(z,\zeta_1,\zeta_2;m_Q,\mu_0)=\frac{1}{N_c^2-1}\frac{1}{2m_Q}\,\delta(1-z)\,\delta(\zeta_1)\,\delta(\zeta_2).
\end{split}
\end{align}
\end{linenomath*}
A complete list of finite LO matching coefficients are given in Appendix~\ref{app:doubleresults}.

\section{NLO matching coefficients}\label{sec:NLO}

The NLO short-distance coefficients in Eq.~(\ref{eq:QQFFNR2}) can be derived by expanding both sides of the factorized formula to NLO as
\begin{linenomath*}
\begin{align}\label{eq:QQFFNRmatchNLO}
\begin{split}
{\cal D}^{\text{NLO}}_{[\cc(\kappa)]\to [\cc(n')]}(z,\zeta_1,\zeta_2;m_Q,\mu_0)
&= \hat{d}^{\text{ NLO}}_{[\cc(\kappa)\to [\cc(n')]}(z,\zeta_1,\zeta_2;m_Q,\mu_0,\mu_\Lambda)\\
&\hspace{-2cm}
+
\sum_{[\cc(n)]} \hat{d}^{\text{ LO}}_{[\cc(\kappa)]\to [\cc(n)]}(z,\zeta_1,\zeta_2;m_Q,\mu_0){{\langle{\cal O}_{[\cc(n)]}^{[\cc(n')]}(\mu_\Lambda)\rangle^{\text{NLO}}}
}.
\end{split}
\end{align}
\end{linenomath*}
If NRQCD factorization is valid to this order, the second term on the RHS should have the same IR divergence as that of the LHS, so that $\hat{d}^{\text{ NLO}}_{[\cc(\kappa)\to [\cc(n')]}(z,\zeta_1,\zeta_2;m_Q,\mu_0,\mu_\Lambda)$ is IR finite.

\begin{figure}[htb]
\begin{center}
\includegraphics[width=0.45\textwidth]{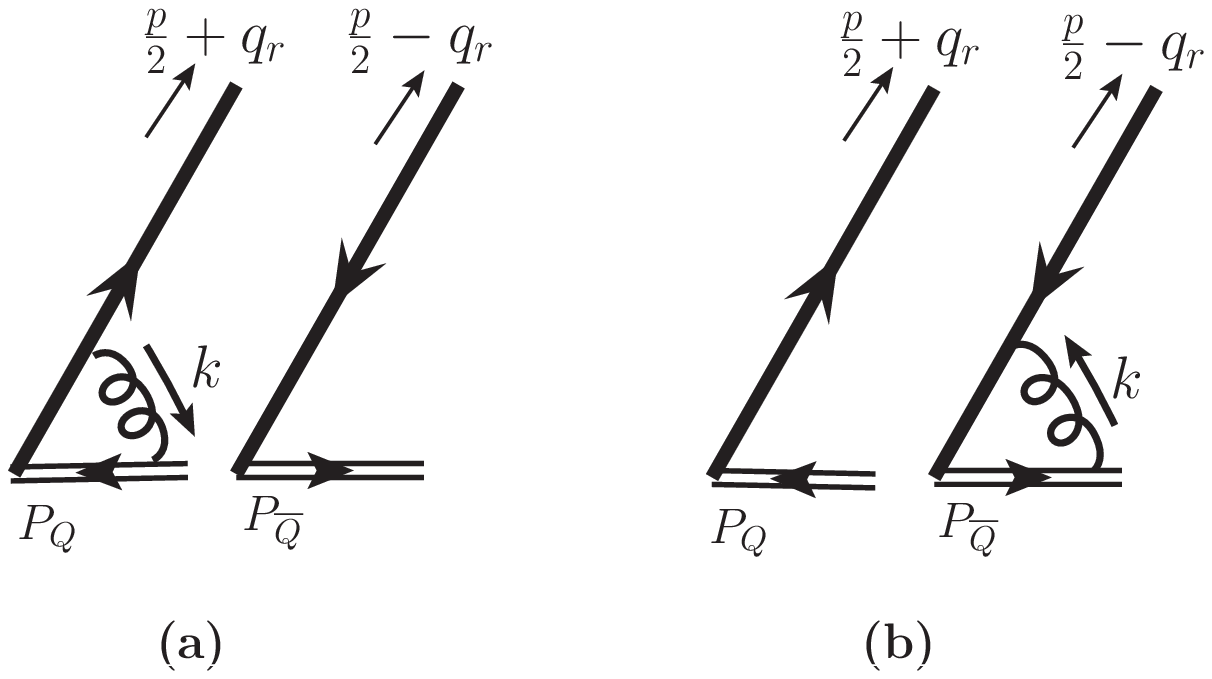}
\includegraphics[width=0.43\textwidth]{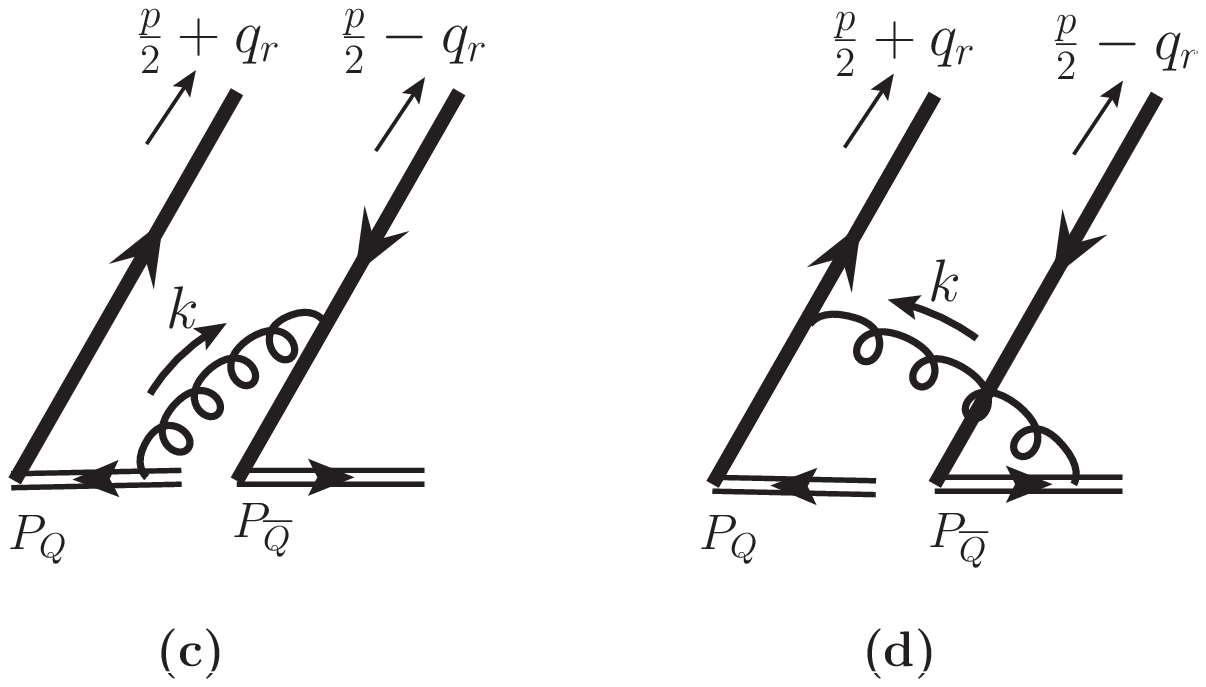}\\
\includegraphics[width=0.19\textwidth]{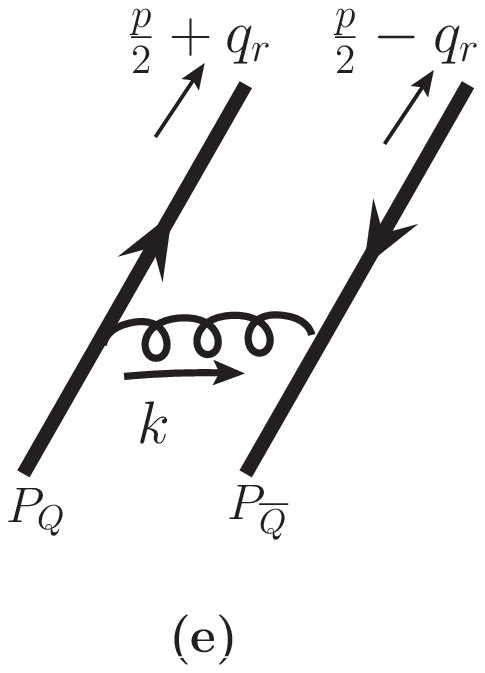}
\includegraphics[width=0.7\textwidth]{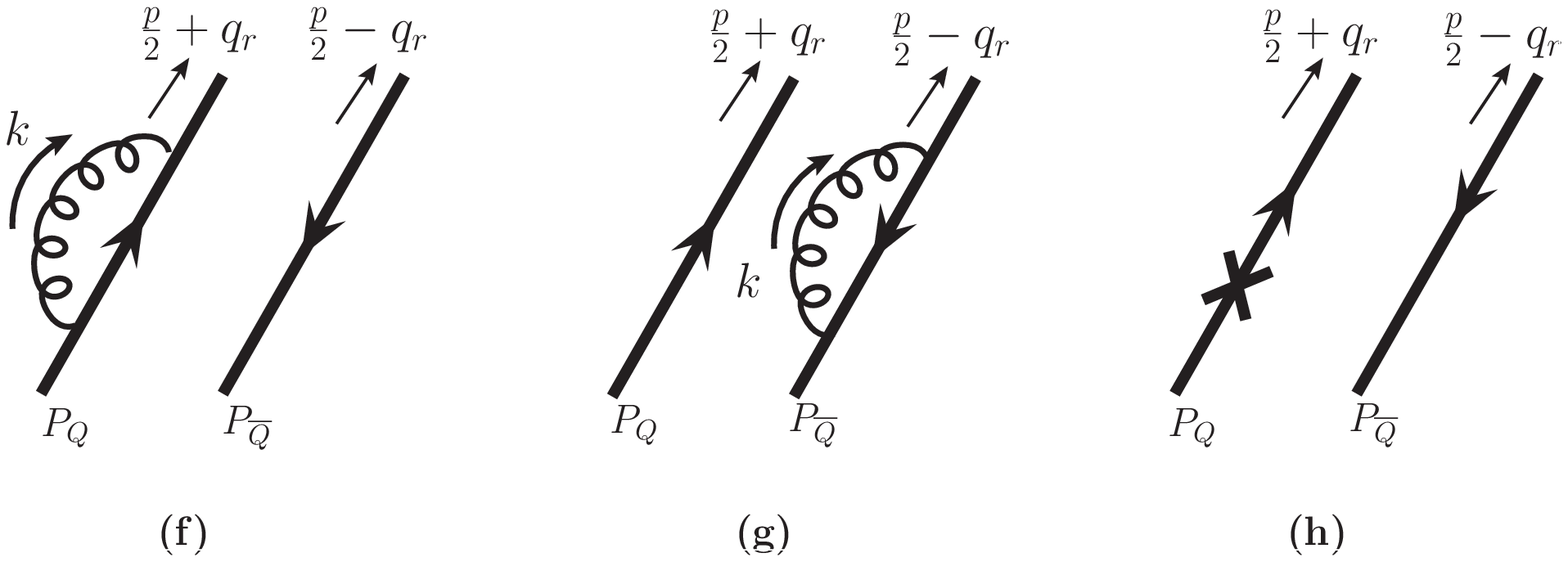}\\
\includegraphics[width=0.19\textwidth]{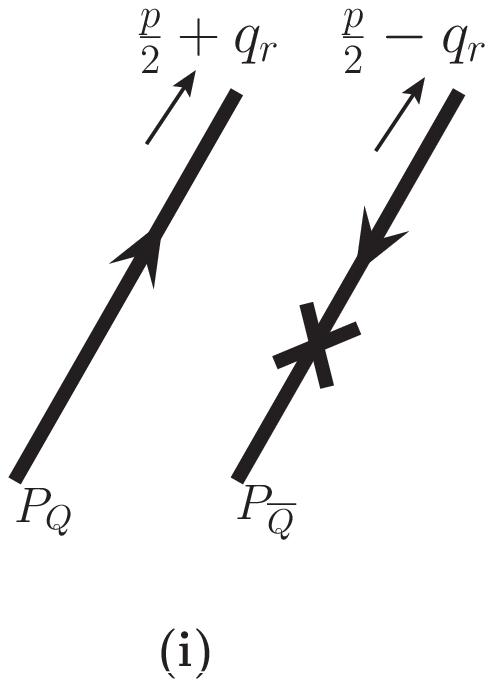}
\hskip 0.2in
\includegraphics[width=0.20\textwidth]{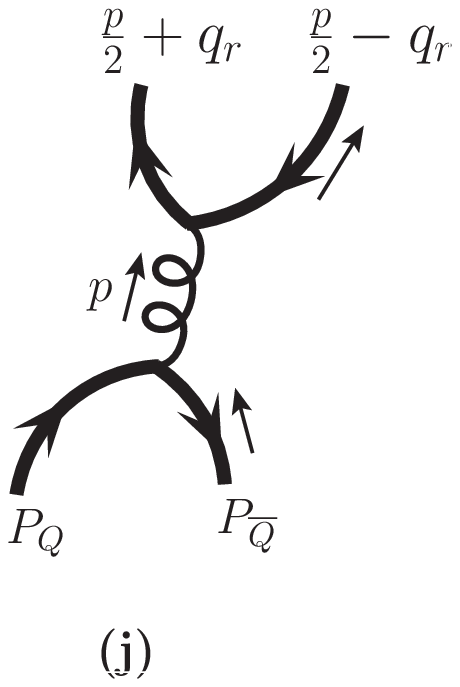}
\hskip 0.15in
\includegraphics[width=0.47\textwidth]{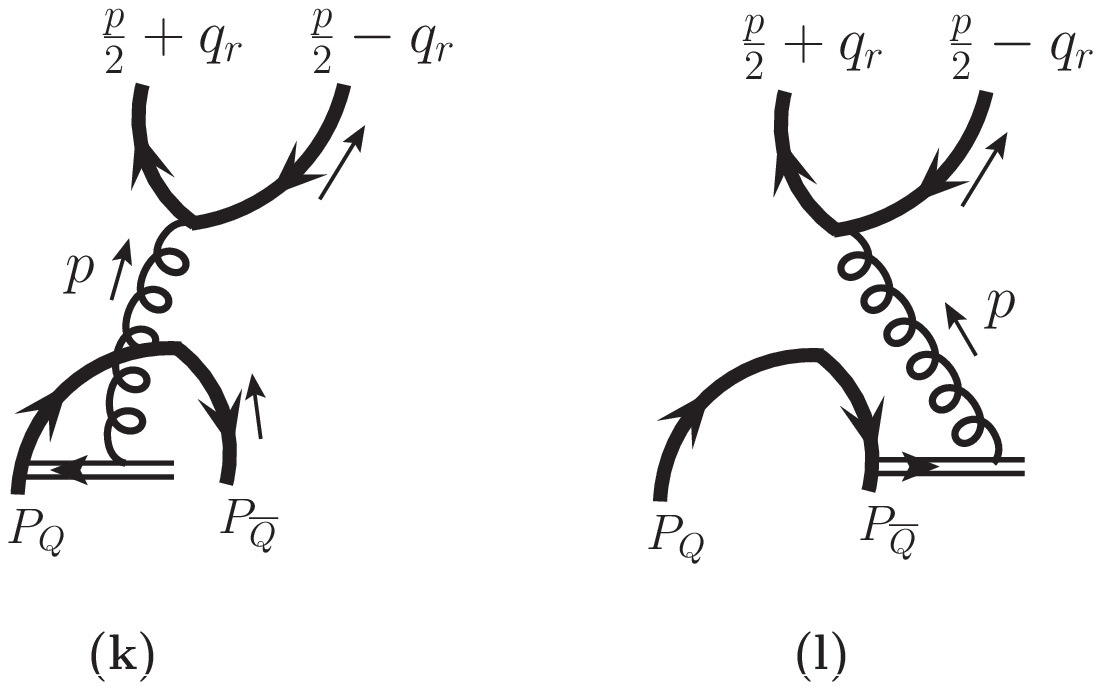}\\
\caption{Feynman diagrams for virtual correction at NLO.
\label{fig:NLOvirtual}}
\end{center}
\end{figure}

\begin{figure}[htb]
\begin{center}
\includegraphics[width=0.45\textwidth]{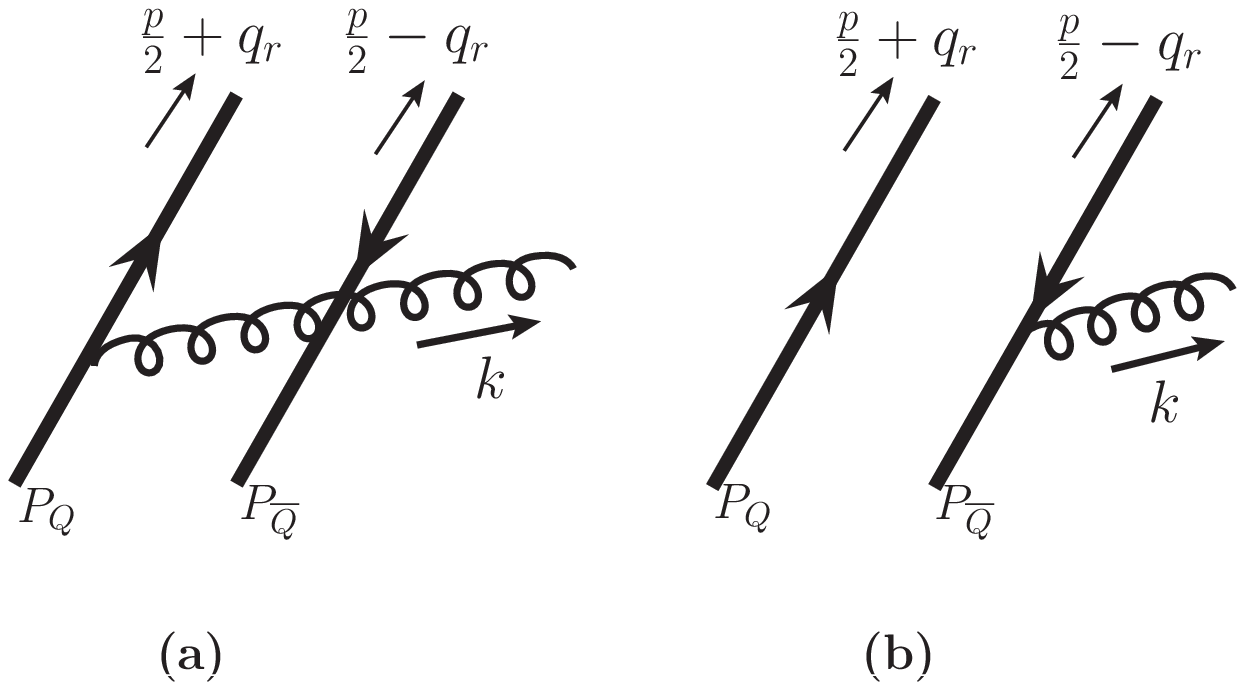}
\includegraphics[width=0.45\textwidth]{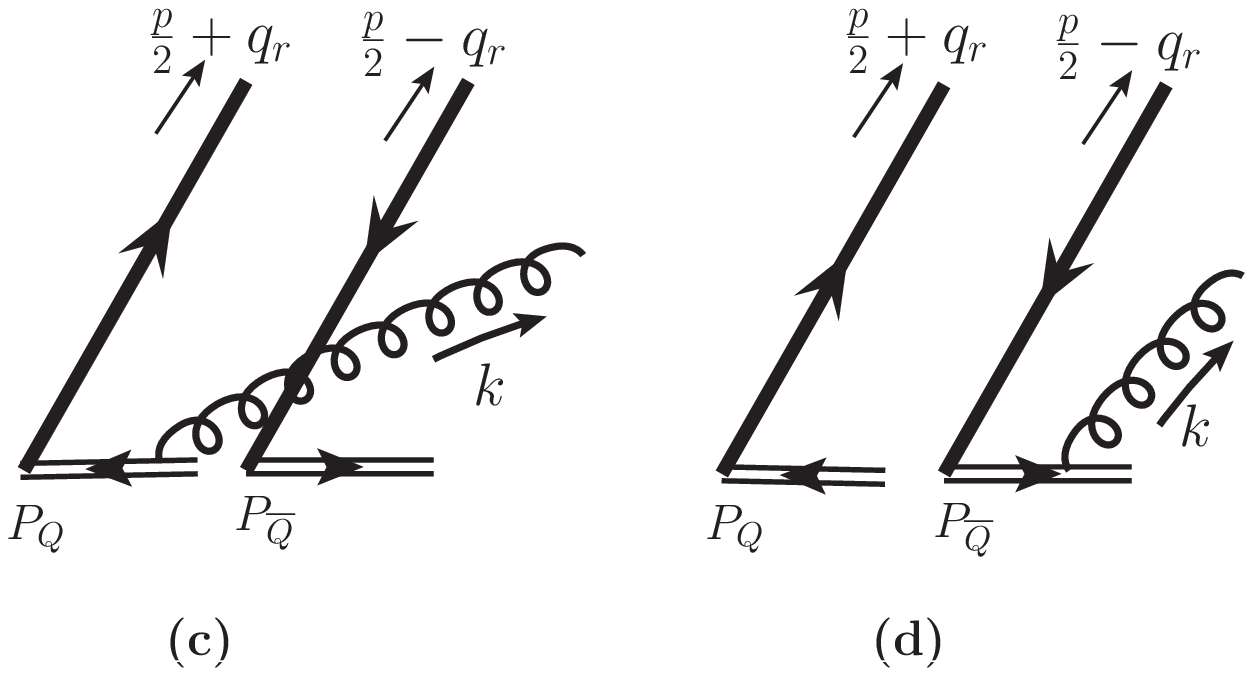}
\caption{Feynman diagrams for real correction at NLO.
\label{fig:NLOreal}}
\end{center}
\end{figure}

${\cal D}^{\text{NLO}}_{[\cc(\kappa)]\to [\cc(n')]}(z,\zeta_1,\zeta_2;m_Q,\mu_0)$ could be calculated directly from Eqs.~(\ref{eq:QQFFQQ})-(\ref{eq:QQFFQQA}) with a proper UV counter term to remove the UV divergence of the composite operators defining the $Q\bar{Q}$-pair FFs.
A general NLO correction includes virtual part and real part. In Feynman gauge, these two parts could be represented in terms of Feynman diagrams in Figs. \ref{fig:NLOvirtual} and \ref{fig:NLOreal}, respectively. Note that the diagrams (c), (d) and (e) in Fig. \ref{fig:NLOvirtual} are loop diagrams, in the sense that they have also imaginary contribution, because of the $q_1$-integral in Eq.~(\ref{eq:QQFFQQA}).

A specific fragmentation process may not have both virtual and real corrections at this order. For example, those which are forbidden at LO do not have NLO virtual correction, while processes with both initial and final $\cc$-pair being in color singlet state do not have NLO real correction. 
In general, fundamental symmetries of QCD not only constrain the structure of the FFs, but also help simplify calculations.  
For example,  the parity and time-reversal invariance of QCD requires the FFs to a unpolarized final-state heavy quarkonium to be symmetric in $\zeta_1$ and $\zeta_2$, which is also consistent with the reality of the FFs.  We also find that a modified charge conjugation, defined as the operation of charge conjugation followed by the replacements of $\zeta_1\to -\zeta_1$ for the scattering amplitude (and $\zeta_2 \to -\zeta_2$ for its complex conjugate) \cite{\Pwave}, is very useful in helping simplify our calculations.  Under this modified charge conjugation, we find that for certain processes, some diagrams can differ from each other only by a global factor, for example, $\pm1$ for virtual correction or some specific $\delta$-function combinations for real correction, or more specifically, only five combinations, as listed in Eq.~(\ref{eq:Delta0}) to (\ref{eq:Delta8}).

We use dimensional regularization to regularize all kinds of divergences in this paper. These divergences include ultra-violet (UV) divergence, infrared (IR) divergence, rapidity divergence and Coulomb divergence. Because of heavy quark mass, there is no collinear divergence. UV divergence of these diagrams will be cancelled by 
the pQCD renormalization of the composite operators defining the $Q\bar{Q}$-pair FFs, where evolution kernels derived in Ref.~\cite{KMQS-hq1} are needed. In general, summation of all diagrams (real and virtual) could still have leftover IR divergence, which should be the same as the IR divergence of LDMEs at NLO.  This must be the case if NRQCD factorization is valid, at least up to this order in $\as$. Rapidity divergence is characterized as $k\cdot \hat{n}\to 0$, with $k$ the momentum of the gluon. Such divergence could overlap with UV divergence and produce a double pole. Eventually, we find the rapidity divergences are cancelled once we sum over all diagrams.

The cancelation of Coulomb divergence needs more discussion. 
The LHS of Eq.~(\ref{eq:QQFFNRmatchNLO}), with the definitions in Eqs.~(\ref{eq:QQFFQQ})-(\ref{eq:QQFFQQA}), is Coulomb divergent
in the region $q_r\sim (m_Qv^2, m_Q \vec{v})$.
Similar terms are also existed in the NLO LDMEs on the RHS of Eq.~(\ref{eq:QQFFNRmatchNLO}).
If NRQCD factorization is valid 
at this order, these terms cancel exactly, leaving the NLO short-distance coefficients free of Coulomb divergence.
However, to show this cancellation, we must keep $q_r$ finite while 
performing the integration of $q_1$, which is usually difficult and tedious.
For our $S$-wave calculations in this paper, we find that  
taking the limit $q_r\to 0$ before doing the $q_1$-integral in Eq.~(\ref{eq:QQFFQQA}) leads to the same result, without explicit Coulomb divergent terms in both the parton-level fragmentation function in the LHS of Eq.~(\ref{eq:QQFFNRmatchNLO}) and in the LDME's  on the RHS of the same equation.
That is, by switching the order of $q_r\to 0$ limit with the $q_1$-integration, one could obtain the same results for NLO short-distance coefficients via a S-wave heavy quark pair, while the Coulomb divergent terms cancel implicitly.  For the $P$-wave case discussed in our companion paper \cite{\Pwave}, we will have to deal with the derivative of $q_r$ in addition to the $q_r\to 0$ limit.  But, as we proved in Ref.~\cite{\Pwave}, the switching between the combination $q_r$-derivative and $q_r\to 0$ limit and the $q_1$-integration in 
Eq.~(\ref{eq:QQFFQQA}) is also valid.

In the rest of this section, we illustrate the detailed NLO calculation with an example:
$[\cc(\aeight)]\to[\cc(\COaSz)]$.
For this channel, the second term on the RHS of Eq.~(\ref{eq:QQFFNRmatchNLO}) vanishes, because for any intermediate state $\cc(n)$, either the LO short-distance coefficient or the NLO LDME is equal to zero (after taking the trick discussed above). Therefore, we have
\begin{linenomath*}
\begin{align}\label{eq:QQFFNRmatchNLOS}
\begin{split}
{\cal D}^{\text{NLO}}_{[\cc(\aeight)]\to[\cc(\COaSz)]}(z,\zeta_1,\zeta_2;m_Q,\mu_0)
&= \hat{d}^{\text{ NLO}}_{[\cc(\aeight)]\to[\cc(\COaSz)]}(z,\zeta_1,\zeta_2;m_Q,\mu_0).
\end{split}
\end{align}
\end{linenomath*}
To calculate the LHS of the above equation, we need to calculate both real and virtual contributions.

\subsection{Real contribution}

The Feynman diagrams for real correction are shown in Fig. \ref{fig:NLOreal}. We calculate these diagrams in both Feynman gauge and light cone gauge, and the results are the same. After some algebra, we derive the real contribution as
\begin{linenomath*}
\begin{align}\label{eq:NLOreal}
\begin{split}
{\cal D}^{\text{NLO}, real}_{[\cc(\aeight)]\to[\cc(\COaSz)]}(z,\zeta_1,\zeta_2;m_Q,\mu_0)
=&
\frac{\alpha_s}{4\pi m_Q N_c (N_c^2-1)}
\left(\frac{4\pi\mu^2}{(2m_Q)^2}\right)^\epsilon \Gamma(1+\epsilon)\\
&\hspace{-3.5cm}
\times
\left\{
-N_c^2\delta(\zeta_1)\delta(\zeta_2)\delta(1-z)\left(\frac{1}{\epsilon_{\text{UV}}\epsilon_{\text{IR}}}-\frac{1}{\epsilon_{\text{IR}}}\right)
+\frac{1}{\epsilon_{\text{UV}}}\frac{z}{(1-z)_+}\frac{\DPlusEight}{4}
\right.\\
&\hspace{-2.8cm}
\left.
+\frac{\DPlusEight}{4}
\left[
-\frac{1}{(1-z)_+}-2\left(\frac{\text{ln}(1-z)}{1-z}\right)_+
+2\,\text{ln}(1-z)+1
\right]
\right\},
\end{split}
\end{align}
\end{linenomath*}
where $(2m_Q)^2=p^2$ in the first line is the invariant mass squared of the produced heavy quark pair, and
\begin{linenomath*}
\begin{align}\label{eq:DPlus8}
\begin{split}
\DPlusEight
\equiv &
\,
4\Big\{
(N_c^2-2)[\delta (1-z+\zeta_1) \delta (1-z+\zeta_2)+\delta (1-z-\zeta_1) \delta (1-z-\zeta_2)]\\
&
+2[\delta (1-z-\zeta_1) \delta (1-z+\zeta_2)+\delta (1-z+\zeta_1) \delta (1-z-\zeta_2)]
\Big\}.
\end{split}
\end{align}
\end{linenomath*}
The origin of each pole is labelled by subscript ``UV" or ``IR". Infrared divergence at $z\to 1$ are extracted with plus prescription
\begin{linenomath*}
\begin{align}\label{eq:zplus}
\begin{split}
\frac{1}{(1-z)^{1+2\epsilon}}
=
-\frac{1}{2\,\epsilon_{\text{IR}}}\delta(1-z)
+\frac{1}{(1-z)_+}
-2\,\epsilon_{\text{IR}}\left(\frac{\text{ln}(1-z)}{1-z}\right)_+.
\end{split}
\end{align}
\end{linenomath*}
The double pole is from the region $k\cdot \hat{n}\to 0, k_\bot \to \infty$. The function is even for both $\zeta_1$ and $\zeta_2$, which is required by charge conjugation symmetry \cite{Pwave}.

In Eq.~(\ref{eq:NLOreal}), the multiplicative factor, $(4\pi\mu^2/p^2)^\epsilon$ with $p^2=(2m_Q)^2$, is a generic feature of one loop calculation using the dimensional regularization, where for the real contribution, $p^2$ is the invariant mass squared of the produced heavy quark pair.  On the other hand, for the virtual contribution, which will be derived in the next subsection, the corresponding multiplicative factor will be
$(4\pi\mu^2/(p/2)^2)^\epsilon=(4\pi\mu^2/m_Q^2)^\epsilon$ with the invariant mass of produced heavy quark or antiquark $(p/2)^2=m_Q^2$.   To prepare for the sum with the virtual correction from the next subsection,
we rewrite the multiplicative factor of the real contribution, $(4\pi\mu^2/(2m_Q)^2)^\epsilon$ as
$(4\pi\mu^2/m_Q^2)^\epsilon\times 4^{-\epsilon}$, so that the real contribution in Eq.(\ref{eq:NLOreal}) can be expressed as
\begin{linenomath*}
\begin{align}\label{eq:NLOreal2}
\begin{split}
{\cal D}^{\text{NLO}, real}_{[\cc(\aeight)]\to[\cc(\COaSz)]}(z,\zeta_1,\zeta_2;m_Q,\mu_0)
=&
\frac{\alpha_s}{4\,\pi m_Q N_c (N_c^2-1)}
\left(\frac{4\pi\mu^2}{m_Q^2}\right)^\epsilon \Gamma(1+\epsilon)\\
&\hspace{-3.5cm}
\times
\left\{
-N_c^2\delta(\zeta_1)\delta(\zeta_2)\delta(1-z)\left(\frac{1}{\epsilon_{\text{UV}}\epsilon_{\text{IR}}}-\frac{1}{\epsilon_{\text{IR}}}\right)
+\frac{1}{\epsilon_{\text{UV}}}\frac{z}{(1-z)_+}\frac{\DPlusEight}{4}
\right.
\\
&\hspace{-2.8cm}
+2\,(\text{ln}\,2)\,N_c^2\,\delta(\zeta_1)\,\delta(\zeta_2)\,\delta(1-z)\frac{1}{\epsilon_{\text{UV}}}
\\
&\hspace{-2.8cm}
-2\left[(\text{ln}\,2)^2+\text{ln}\,2\right]N_c^2\,\delta(\zeta_1)\,\delta(\zeta_2)\,\delta(1-z)
-(2\, \text{ln}\, 2)\frac{z}{(1-z)_+}\frac{\DPlusEight}{4}
\\
&\hspace{-2.8cm}
\left.
+\frac{\DPlusEight}{4}
\left[
-\frac{z}{(1-z)_+}-2\,z\,\left(\frac{\text{ln}(1-z)}{1-z}\right)_+
\right]
\right\},
\end{split}
\end{align}
\end{linenomath*}
where terms with $\ln2$ dependence are due to the multiplication of the $4^{-\epsilon}$ with the poles, and terms vanishing at $D=4$ are neglected. Note that since the $4^{-\epsilon}$ originates from infra-red region, its ${\cal O}(\epsilon)$ term should first cancel with $1/\epsilon_{\text{IR}}$ pole
before it cancels the $1/\epsilon_{\text{UV}}$ pole.
The mismatch between $p^2$ of the real contribution and the $(p/2)^2$ of the virtual contribution is
similar to the phase space mismatch between the real and virtual contribution to the evolution kernels
of heavy quark fragmentation functions, which led to the $\ln(u\bar{u}v\bar{v})$ term in the kernels
\cite{KMQS-hq1}.
Actually, such mismatch was originated from the difference of the gluon's maximum allowed light-cone momentum between the real and the virtual diagrams \cite{KMQS-hq1}.

\subsection{Virtual contribution}

In Feynman gauge, Feynman diagrams for virtual correction are shown in Fig. \ref{fig:NLOvirtual}. Note that diagrams (j), (k) and (l) in Fig. \ref{fig:NLOvirtual} have no contributions for ${[\cc(\aeight)]\to[\cc(\COaSz)]}$ kernel. The full virtual contribution could be thus written as
\begin{linenomath*}
\begin{align}\label{eq:NLOvirtual}
\begin{split}
&
\hspace{-0.25cm}
{\cal D}^{\text{NLO}, virtual}_{[\cc(\aeight)]\to[\cc(\COaSz)]}(z,\zeta_1,\zeta_2;m_Q,\mu_0)
=
2\delta(1-z)\delta(\zeta_2)\left\{\Lambda(\zeta_1)+\Sigma(\zeta_1)+\Pi(\zeta_1)+W(\zeta_1)\right\}
\\
&\hspace{4cm}
+2
\delta(1-z)\delta(\zeta_1)\left\{\Lambda^\dag(\zeta_2)+\Sigma^\dag(\zeta_2)+\Pi^\dag(\zeta_2)+W^\dag(\zeta_2)\right\},
\end{split}
\end{align}
\end{linenomath*}
where the first (second) line is from the cut-notation diagrams with NLO diagrams in Fig. \ref{fig:NLOvirtual} on the left (right) and LO diagrams on the right (left). Each line is further separated into four terms corresponding to different diagrams in Fig. \ref{fig:NLOvirtual}: $\Lambda$ for diagrams (a) and (b), $\Sigma$ for diagrams (c) and (d), $\Pi$ for diagram (e), and $W$ for diagrams (f), (g), (h) and (i). $\zeta_1$ and $\zeta_2$ could be any number between $-1$ and $1$. From charge conjugation symmetry \cite{Pwave}, we find that ${\cal D}^{\text{NLO}, virtual}_{[\cc(\aeight)]\to[\cc(\COaSz)]}(z,\zeta_1,\zeta_2;m_Q,\mu_0)$ is an even function of $\zeta_1$ and $\zeta_2$, and therefore, $\Lambda(\zeta_1)$, $\Sigma(\zeta_1)$, $\Pi(\zeta_1)$ and $W(\zeta_1)$ must be even functions of $\zeta_1$ for this process, which is manifested in our results below.

In the calculation of this virtual contribution,  we encounter $\zeta_1^{-1-2\epsilon}\,\text{Sgn}(\zeta_1)$ and $\zeta_1^{-2-2\epsilon}$ type terms, which are divergent as $\zeta_1\to 0$. We regularize these singular terms with a set of generalized plus-distributions
\begin{linenomath*}
\begin{subequations}\label{eq:plus12}
\begin{align}
&\frac{\text{Sgn}(\zeta_1)}{(\zeta_1)^{1+2\epsilon}}
=
-\frac{1}{\epsilon_{\text{IR}}}\delta(\zeta_1)
+\left(\frac{1}{\zeta_1}\right)_{1+}
-\epsilon\left(\frac{\text{ln}(\zeta_1^2)}{\zeta_1}\right)_{1+},\label{eq:zeta1}\\
&\frac{1}{(\zeta_1)^{2+2\epsilon}}
=
-2(1-2\epsilon)\delta(\zeta_1)
+\left(\frac{1}{\zeta_1^2}\right)_{2+}
-\epsilon\left(\frac{\text{ln}(\zeta_1^2)}{\zeta_1^2}\right)_{2+}.\label{eq:zeta2}
\end{align}
\end{subequations}
\end{linenomath*}
In the same manner, we also define
\begin{linenomath*}
\begin{align}
&\text{Sgn}(\zeta_1){\,\zeta_1}
=
(\zeta_1)_{0+}.
\end{align}
\end{linenomath*}
%
These generalized plus-distributions are collectively defined as
\begin{linenomath*}
\begin{align}\label{eq:plusminusalldef}
\begin{split}
\Big(g(\zeta_1)\Big)_{m+}
\equiv
\int_{-1}^{1}\left[\theta(x) + \theta(-x)\right] g(|x|)
\times
\left(\delta(x-\zeta_1)-\sum_{i=0}^{m-1}\frac{\delta^{(i)}(\zeta_1)}{i\,!}(-x)^{\,i}\right)dx\, ,
\end{split}
\end{align}
\end{linenomath*}
where $m$ is a non-negative integer.  These new plus-distributions are
even functions of $\zeta_1$.
For any well-behaved function, $f(\zeta_1)$, which, so as its derivatives up to the $(m-1)$-th order,
should be free of divergence for $\zeta_1 \in [-1,1]$, we have
\begin{linenomath*}
\begin{align}\label{eq:plusminusall}
\begin{split}
\int \Big(g(\zeta_1)\Big)_{m+}f(\zeta_1)\,d\zeta_1
=
\int_{-1}^{1}\left[\theta(\zeta_1) + \theta(-\zeta_1)\right] g(|\zeta_1|)
\times
\left(f(\zeta_1)-\sum_{i=0}^{m-1}\frac{f^{(i)}(0)}{i\,!}\zeta_1^{\,i}\right)d\zeta_1,
\end{split}
\end{align}
\end{linenomath*}
where $f^{(i)}(\zeta_1)$ is the $i$th derivative of $f(\zeta_1)$.

After considerable amount of algebra, we derive the four terms contributing to
the virtual contribution in Eq.~(\ref{eq:NLOvirtual}),
\begin{linenomath*}
\begin{subequations}
\begin{align}
\Lambda(\zeta_1)
=&
\frac{\alpha_s C_F}{8\pi m_Q (N_c^2-1)}\delta(\zeta_1)
\left(\frac{4\pi\mu^2}{m_Q^2}\right)^\epsilon\Gamma(1+\epsilon)
\left(\frac{1}{\epsilon_{\text{UV}}\epsilon_{\text{IR}}}+\frac{2}{\epsilon_{\text{UV}}}+4\right),\\
\Sigma(\zeta_1)
=&
\frac{\alpha_s}{8\pi m_Q}\frac{1}{2N_c (N_c^2-1)}
\left(\frac{4\pi\mu^2}{m_Q^2}\right)^\epsilon\Gamma(1+\epsilon)
\left\{\frac{1}{\epsilon_{\text{UV}}\epsilon_{\text{IR}}}\delta(\zeta_1)+\frac{1}{\epsilon_{\text{UV}}}\left[\Plusz{1}-\left(\frac{1}{\zeta_1}\right)_{1+}\right]\right.\nonumber
\\
&\hspace{4cm}\left.
+\left(\frac{\text{ln}(\zeta_1^2)}{\zeta_1}\right)_{1+}-\Plusz{\text{ln}(\zeta_1^2)}\right\},
\\
\Pi(\zeta_1)
=&
\frac{\alpha_s}{16\pi m_Q}\frac{1}{2N_c(N_c^2-1)}
\left(\frac{4\pi\mu^2}{m_Q^2}\right)^\epsilon\Gamma(1+\epsilon)
\left\{\frac{1}{\epsilon_{\text{UV}}}\Big[(\zeta_1)_{0+}-\Plusz{1}\Big]-\frac{2}{\epsilon_{\text{IR}}}\delta(\zeta_1)+4\,\delta(\zeta_1)\right.\nonumber
\\
&\hspace{1.5cm}\left.
-\Big(\zeta_1\,\text{ln}(\zeta_1^2)+\zeta_1\Big)_{0+}+2\left(\frac{1}{\zeta_1}\right)_{1+}-2\left(\frac{1}{\zeta_1^2}\right)_{2+}+\Plusz{\text{ln}(\zeta_1^2)+1}\right\},
\\
W(\zeta_1)
=&
-\frac{\alpha_s C_F}{16\pi m_Q(N_c^2-1)}\delta(\zeta_1)
\left(\frac{4\pi\mu^2}{m_Q^2}\right)^\epsilon\Gamma(1+\epsilon)
\left(\frac{1}{\epsilon_{\text{UV}}}+\frac{2}{\epsilon_{\text{IR}}}+4\right),
\end{align}
\end{subequations}
\end{linenomath*}
where $C_F=(N_c^2-1)/(2N_c)$. It is straightforward to verify that every function above is even for $\zeta_1$. Substitute these expressions into Eq.~(\ref{eq:NLOvirtual}). we obtain the NLO virtual correction
\begin{linenomath*}
\begin{align}\label{eq:NLOvirtual2}
\begin{split}
\hspace{1cm}&\hspace{-1cm}{\cal D}^{\text{NLO}, virtual}_{[\cc(\aeight)]\to[\cc(\COaSz)]}(z,\zeta_1,\zeta_2;m_Q,\mu_0)
=
\frac{\alpha_s}{8\pi m_Q}\frac{1}{2N_c (N_c^2-1)}\delta(1-z)\delta(\zeta_2)
\left(\frac{4\pi\mu^2}{m_Q^2}\right)^\epsilon\Gamma(1+\epsilon)\\
&
\times
\left\{
2\,N_c^2\delta(\zeta_1)\left[\frac{1}{\epsilon_{\text{UV}}\epsilon_{\text{IR}}}-\frac{1}{\epsilon_{\text{IR}}}+2\right]
+\frac{1}{\epsilon_{\text{UV}}}\left[3(N_c^2-1)\delta(\zeta_1)-2\left(\frac{1}{\zeta_1}\right)_{1+}+\Plusz{\zeta_1+1}\right]
\right.
\\
&
\left.
+2\left[\left(\frac{1}{\zeta_1}\right)_{1+}-\left(\frac{1}{\zeta_1^2}\right)_{2+}+\left(\frac{\text{ln}(\zeta_1^2)}{\zeta_1}\right)_{1+}\right]-\Plusz{(\zeta_1+1)\text{ln}(\zeta_1^2)}-\Plusz{\zeta_1-1}
\right\}
\\
&
+(\zeta_1\longleftrightarrow \zeta_2).
\end{split}
\end{align}
\end{linenomath*}
We also derive the same result by using the light-cone gauge.  As noted in the last subsection,
there is a mismatch between the $\left({4\pi \mu^2}/{(2m_Q)^2}\right)^\epsilon$ of the real correction
in Eq.~(\ref{eq:NLOreal}) and the $\left({4\pi \mu^2}/{m_Q^2}\right)^\epsilon$ of the virtual correction
in Eq.~(\ref{eq:NLOvirtual2}), which led to the extra logarithms in Eq.~(\ref{eq:NLOreal2}).

Comparing Eq.~(\ref{eq:NLOreal2}) and Eq.~(\ref{eq:NLOvirtual2}), all infrared poles cancel between the real and virtual contributions. However, the sum of Eq.~(\ref{eq:NLOreal2}) and Eq.~(\ref{eq:NLOvirtual2}) still contains ultraviolet divergence, which should be taken care of by the renormalization of the nonlocal operators defining the fragmentation functions in Eq.~(\ref{eq:QQFF}).

\subsection{Renormalization of composite operators defining FFs}

As defined in Eq.~(\ref{eq:QQFFQQ}), the heavy quark-pair FF is defined with a UV counter-term,
which is a result of the UV renormalization of the composite operators defining the FFs.
The UV counter-term removes the perturbative UV divergence of the FFs order-by-order in powers
of $\alpha_s$.
In general, UV divergence of FFs calculated by using the NRQCD factorization should be different from that defined by pQCD factorization. The reason is following. The heavy quark mass in pQCD factorization is a small scale and is set to be zero at the beginning, while the heavy quark mass in NRQCD factorization is a large scale and is always kept to be finite. Because of the finite quark mass, there are helicity flip contribution in NRQCD calculation, which is forbidden in pQCD factorization. Therefore, extra UV divergence for helicity flip contribution may emerge in NRQCD calculation. An example for this kind of UV divergence is the contribution of the diagram (j) in Fig.~\ref{fig:NLOvirtual} in the NLO calculation of ${\cal D}_{[\cc(t^{[8]})]\to [\cc(\COcSa)]}$. Thus, the correct way to renormalize input FFs calculated by using the NRQCD factorization is:
\begin{linenomath*}
\begin{align}\label{eq:gren0}
{\cal D}_{[Q\bar{Q}(\kappa)]\to[Q\bar{Q}(n)]}=\Gamma_{[Q\bar{Q}(\kappa)]\to[Q\bar{Q}(\kappa')]}
\otimes
{\cal Z}_{[Q\bar{Q}(\kappa')]\to[Q\bar{Q}(\kappa'')]}
\otimes
{\cal D}_{[Q\bar{Q}(\kappa'')]\to[Q\bar{Q}(n)]}^{bare},
\end{align}
\end{linenomath*}
where $\Gamma$ is the evolution kernel defined in pQCD factorization, and $\cal Z$ is used to take care of the extra UV divergence discussed above.
Expanding Eq.~(\ref{eq:gren0}) to NLO, we find
\begin{linenomath*}
\begin{align}\label{eq:grenNLO}
{\cal D}_{[Q\bar{Q}(\kappa)]\to[Q\bar{Q}(n)]}^{\text{NLO}}={\cal D}_{[Q\bar{Q}(\kappa)]\to[Q\bar{Q}(n)]}^{\text{\,NLO},bare}
+{\cal D}_{[Q\bar{Q}(\kappa)]\to[Q\bar{Q}(n)]}^{\text{\,NLO},ren, \text{1}}
+{\cal D}_{[Q\bar{Q}(\kappa)]\to[Q\bar{Q}(n)]}^{\text{\,NLO},ren, \text{2}},
\end{align}
\end{linenomath*}
where
\begin{linenomath*}
\begin{align}\label{eq:gren1}
\begin{split}
{\cal D}^{\text{NLO}, ren,1}_{[Q\bar{Q}(\kappa)]\to[Q\bar{Q}(n)]}
&(z,\zeta_1,\zeta_2;m_Q)
=
{\Gamma}_{[Q\bar{Q}(\kappa)]\to[Q\bar{Q}(\kappa')]}^{\text{NLO}}
\otimes
{\cal D}^{\text{ LO}, bare}_{[Q\bar{Q}(\kappa')]\to[Q\bar{Q}(n)]}
\\
&\hspace{-1cm}
=
-
\frac{A}{\epsilon_{\text{UV}}}\,
\sum_{[\cc(\kappa')]}\int_z^1\frac{dz'}{z'}\int_{-1}^1 \frac{d\zeta'_1\, d\zeta'_2}{4} \,
\\
&\hspace{-1cm}
\times
\Gamma_{[\cc(\kappa)]\to [\cc(\kappa')]}(z',u=\frac{1+\zeta_1}{2},v=\frac{1+\zeta_2}{2};u'=\frac{1+\zeta'_1}{2},v'=\frac{1+\zeta'_2}{2}) \,
\\
&\hspace{-1cm}
\times
{\cal D}^{\text{ LO}, bare}_{[Q\bar{Q}(\kappa')]\to[Q\bar{Q}(n)]}(z/z',\zeta'_1,\zeta'_2,m_Q)\, ,
\end{split}
\end{align}
\end{linenomath*}
the summation runs over all possible perturbative intermediate $\cc$-pair states $\kappa'$, and
${\Gamma}_{[Q\bar{Q}(\kappa)]\to[Q\bar{Q}(\kappa')]}$ is the evolution kernel for a heavy quark pair to evolve into another heavy quark pair perturbatively, which is process-independent and has been derived in Refs.~\cite{Fleming:2013qu,KMQS-hq1}. In this paper, we will use results obtained in Ref.~\cite{KMQS-hq1}.  In Eq.~(\ref{eq:gren1}), LO short-distance coefficient ${\cal D}^{\text{ LO}, bare}_{[\cc(\kappa')]\to[\cc(n)]}={\cal D}^{\text{ LO}}_{[\cc(\kappa')]\to[\cc(n)]}$ could be similarly derived as the example in the last section, but, must be evaluated and kept in $D$-dimension in the dimensional regularization.  The proportional factor, $A=1+{\cal O}(\epsilon)$ in Eq.~(\ref{eq:gren1}), is a constant whose choice determines the renormalization scheme.

${\cal D}_{[Q\bar{Q}(\kappa)]\to[Q\bar{Q}(n)]}^{\text{\,NLO},ren, \text{2}}$ in Eq.~(\ref{eq:grenNLO}) is defined as
\begin{linenomath*}
\begin{align}\label{eq:gren2}
{\cal D}_{[Q\bar{Q}(\kappa)]\to[Q\bar{Q}(n)]}^{\text{\,NLO},ren, \text{2}}={\cal Z}_{[Q\bar{Q}(\kappa)]\to[Q\bar{Q}(\kappa')]}^{\text{NLO}}
\otimes
{\cal D}^{\text{ LO}, bare}_{[Q\bar{Q}(\kappa')]\to[Q\bar{Q}(n)]},
\end{align}
\end{linenomath*}
and can be similarly written in the integration form as Eq.~(\ref{eq:gren1}). ${\cal D}_{[Q\bar{Q}(\kappa)]\to[Q\bar{Q}(n)]}^{\text{\,NLO},ren, \text{2}}$ is scheme dependent, and we will use the same scheme as that in Eq.~(\ref{eq:gren1}). Then, in our calculation, ${\cal D}_{[Q\bar{Q}(\kappa')]\to[Q\bar{Q}(n)]}^{\text{\,NLO},ren, \text{2}}$ equals to zero for all processes except $[\cc(t^{[8]})]\to[\cc(\COcSa)]$, where
\begin{linenomath*}
\begin{align}\label{eq:gren4}
{\cal D}_{[Q\bar{Q}(t^{[8]})]\to[Q\bar{Q}(\COcSa)]}^{\, \text{NLO}, ren, 2} =
\frac{\alpha_s\,\delta(1-z)}{48\,\pi\, m_Q (N_c^2-1)}\frac{A}{\epsilon_{UV}}
\big[ \delta(\zeta_1)\Plusz{1}+\delta(\zeta_2)\Plusz{1} \big].
\end{align}
\end{linenomath*}
In the following, we use ${\cal D}_{[Q\bar{Q}(\kappa)]\to[Q\bar{Q}(n)]}^{\text{\,NLO},ren}$ to represent the addition of the two counter-terms in Eq~(\ref{eq:grenNLO}).

For our example $[\cc(\aeight)]\to[\cc(\COaSz)]$, from Eq.~(\ref{eq:gren1}), one could conclude that the LO short-distance coefficients vanish unless $[\cc(\kappa)]$ is $\aeight$.  Therefore, Eq.~(\ref{eq:gren1}) could be reduced to
\begin{linenomath*}
\begin{align}\label{eq:ren1}
\begin{split}
{\cal D}^{\text{NLO}, ren}_{[\cc(\aeight)]\to[\cc(\COaSz)]}(z,\zeta_1,\zeta_2;m_Q)
=&
-
\frac{A}{\epsilon_{\text{UV}}}\frac{1}{8 m_Q(N_c^2-1)}
\\
&\hspace{-3cm}
\times
\Gamma_{[\cc(a^{[8]})]\to [\cc(a^{(8)})]}(z,u=\frac{1+\zeta_1}{2},v=\frac{1+\zeta_2}{2};u'=\frac{1}{2},v'=\frac{1}{2}),
\end{split}
\end{align}
\end{linenomath*}
where we have used the result of LO short-distance coefficient in Eq.~(\ref{eq:resLOS}) and performed the integration with the $\delta$-functions.  The evolution kernel has been calculated in Ref.~\cite{KMQS-hq1} as
\begin{linenomath*}
\begin{align}\label{eq:kernelconv}
\begin{split}
&\hspace{-0.3cm}
\Gamma_{[\cc(a^{[8]})]\to [\cc(a^{(8)})]}(z,u=\frac{1+\zeta_1}{2},v=\frac{1+\zeta_2}{2};u'=\frac{1}{2},v'=\frac{1}{2}),
\\
&
\hspace{-0.5cm}
=
\left(\frac{\alpha_s}{2\pi}\right)
\frac{1}{N_c}
\Big\{
\frac{z}{(1-z)_+}\DPlusEight
+8\,(\text{ln}2)\,N_c^2\,\delta(\zeta_1)\,\delta(\zeta_2)\delta(1-z)
+\delta(1-z)[\delta(\zeta_2)F(\zeta_1)+\delta(\zeta_1)F(\zeta_2)]
\Big\},
\end{split}
\end{align}
\end{linenomath*}
where $\DPlusEight$ is given by Eq.~(\ref{eq:DPlus8}) and
\begin{linenomath*}
\begin{align}
\begin{split}
F(\zeta_1)\equiv 3\,(N_c^2-1)\, \delta(\zeta_1)-2\left(\frac{1}{\zeta_1}\right)_{1+}+\Plusz{\zeta_1+1}.
\end{split}
\end{align}
\end{linenomath*}
Substitute Eq.~(\ref{eq:kernelconv}) into Eq.~(\ref{eq:ren1}),
we obtain the contribution to the UV counter-term as
\begin{linenomath*}
\begin{align}\label{eq:ren2}
\begin{split}
&{\cal D}^{\text{NLO}, ren}_{[\cc(\aeight)]\to[\cc(\COaSz)]}(z,\zeta_1,\zeta_2;m_Q)
=
-\frac{\alpha_s}{\pi}(4\pi e^{-\gamma_E})^\epsilon\frac{1}{\epsilon_{\text{UV}}}\frac{1}{16\,m_Q N_c(N_c^2-1)}
\\
&\hspace{4cm}
\times
\Big\{
\frac{z}{(1-z)_+}{\DPlusEight}
+8 (\text{ln}\, 2)\,N_c^2 \,\delta(\zeta_1)\,\delta(\zeta_2)\,\delta(1-z)
\\
&\hspace{5cm}
+\delta(1-z)\big[\delta(\zeta_2)F(\zeta_1)+\delta(\zeta_1)F(\zeta_2)\big]
\Big\},
\end{split}
\end{align}
\end{linenomath*}
where the ``$A$'' in Eq.~(\ref{eq:ren1}) was chosen to be $(4\pi e^{-\gamma_E})^\epsilon$ for $\overline{\rm MS}$ scheme.  It is straightforward to verify the cancellation of UV divergence by adding up Eqs.~(\ref{eq:NLOreal2}), (\ref{eq:NLOvirtual2}) and (\ref{eq:ren2}).

From Eq.~(\ref{eq:QQFFNRmatchNLOS}), we obtain the NLO short-distance coefficient,
\begin{linenomath*}
\begin{align}\label{eq:resNLOS}
\begin{split}
{\hat d}^{\text{ NLO}}_{[\cc(a^{[8]})]\to [\cc(^1S_0^{[8]})]}(z,\zeta_1,\zeta_2;m_Q,\mu_0)
=&
\frac{\alpha_s}{16\,\pi m_Q(N_c^2-1)}
\\
&\hspace{-3cm}
\times
\Big\{
\left(\frac{2\pi}{\alpha_s}\right)
\Gamma_{[\cc(\aeight)]\to [\cc(\aeight)]}(z,\frac{1+\zeta_1}{2},\frac{1+\zeta_2}{2};\frac{1}{2},\frac{1}{2})\, \LogUV
\\
&\hspace{-2cm}
+
R(z,\zeta_1,\zeta_2)
+\delta(1-z)[V(\zeta_1)\delta(\zeta_2)+V(\zeta_2)\delta(\zeta_1)]
\Big\},
\end{split}
\end{align}
\end{linenomath*}
where $R$ and $V$ are finite contributions from real and virtual diagrams, respectively, which are defined as
\begin{linenomath*}
\begin{subequations}\label{eq:identity}
\begin{align}
&
V(\zeta_1)
=
\frac{1}{N_c}
\left\{
2\left[-\left(\frac{1}{\zeta_1^2}\right)_{2+}+\left(\frac{1}{\zeta_1}\right)_{1+}+\left(\frac{\text{ln}(\zeta_1^2)}{\zeta_1}\right)_{1+}\right]-\Plusz{(\zeta_1+1)\text{ln}(\zeta_1^2)}-\Plusz{\zeta_1-1}
\right.\nonumber
\\
&\hspace{2.5cm}
+4N_c^2\delta(\zeta_1)
\bigg\},
\\
&
R(z,\zeta_1,\zeta_2)
=
\frac{1}{N_c}
\bigg\{
{\DPlusEight}
\left[
-2z\left(\frac{\text{ln}(2-2z)}{1-z}\right)_+ -\frac{z}{(1-z)_+}
\right]
\nonumber
\\
&
\hspace{5cm}
-8\left[(\text{ln}\,2)^2+\text{ln}\,2\right]N_c^2\,\delta(\zeta_1)\,\delta(\zeta_2)\,\delta(1-z)
\bigg\},
\end{align}
\end{subequations}
\end{linenomath*}
where $\Delta_-^{(8)}$ is defined in Eq.~(\ref{eq:DPlus8}).
Although this expression is not in the same compact form as what is shown in
Appendix~\ref{app:doubleresults}, it is trivial to verify the equivalence.

We found that all NLO short-distance coefficients for heavy quarkonium FFs from a perturbatively
produced heavy quark pair, calculated in NRQCD factorization formulism, are IR safe.
A complete list of our results are given in Appendix~\ref{app:doubleresults}.


\section{Summary and conclusion}
\label{sec:summary}

We calculated heavy quarkonium FFs at the input scale, $\mu_0\gtrsim 2m_Q$, in terms of NRQCD factorization approach up to ${\cal O}(v^4)$ in its velocity expansion.
We calculated all short-distance coefficients at LO and NLO in powers of $\alpha_s$ to single parton FFs, as well as contributions at the first non-trivial order in $\as$ to the heavy quark pair FFs.  In this paper, we presented detailed calculations of heavy quark pair FFs to a quarkonium
through a $S$-wave non-relativistic $\cc$-pair.
All contributions through a $P$-wave non-relativistic $\cc$-pair are presented in a companion
paper~\cite{\Pwave}.  Although there is no formal proof of NRQCD factorization approach to the FFs,
we found that all perturbative infrared sensitivities are cancelled at the order of our calculations,
which ensures all calculated short-distance coefficients infrared safe.

The new perturbative QCD factorization formalism for evaluating the heavy quarkonium production at large $p_T$ effectively factorizes the production into three stages: (1) perturbative hard partonic collisions at a distance scale of ${\cal O}(1/p_T)$, (2) perturbative resummation of leading logarithmic contributions from the distance scale of ${\cal O}(1/p_T)$ to ${\cal O}(1/\mu_0)$, and (3) nonperturbative dynamics beyond the distance scale of ${\cal O}(1/\mu_0)$, which are covered by the universal FFs at the input scale $\mu_0$.  Since both the physics at stages (1) and (2) are effectively perturbative and independent of heavy quark mass, flavor and spin, they are the same regardless which heavy quarkonium (J/$\psi$, $\psi'$, $\chi_c$, or anyone from the $\Upsilon$ family) is actually produced.  In this QCD factorization formalism, it is these input FFs that are responsible for the characteristics of producing different heavy quarkonium states, including the spin and polarization of the quarkonium produced.
That is, it is the dynamics at the stage (3) at the input scale $\mu_0$ and below that is really responsible for the formation of a heavy quarkonium from a perturbatively produced heavy quark pair.  By applying NRQCD factorization to the input FFs, we effectively further factorize the dynamics at the stage (3) into two:  new-type perturbative physics between the momentum scale of  ${\cal O}(\mu_0)$ and  ${\cal O}(\mu_{\Lambda}) \sim
{\cal O}(m_Q)$, and the nonperturbative physics at the momentum scale $O(m_Q v)$ and below.  If NRQCD factorization is valid for evaluating these FFs, we should be able to systematically predict the production characteristics of all heavy quarkonium states, in terms of limited and universal NRQCD LDMEs.  That is, the proof or disproof of NRQCD factorization for evaluating these FFs is critically important for understanding the heavy quarkonium production, which is still puzzling us after almost forty years since the discovery of $J/\psi$ \cite{Aubert:1974js, Augustin:1974xw}.
Our effort in this and its companion paper is the first step to focus the heavy quarkonium production in terms of the input FFs.  Our results on input FFs bridge the gap between the perturbative QCD factorization formalism and its phenomenological applications.

\section*{Acknowledgments}

We thank Z.-B.~Kang and G.~Sterman for useful discussions and
G.T.~Bodwin and E.~Braaten for helpful communications.
We also thank D. Yang and X. Wang for lots of communications 
for cross-checking the results in the color singlet channel.
This work was supported in part by the U. S. Department of Energy
under contract No. DE-AC02-98CH10886,
and the National Science Foundation under
grant Nos.~PHY-0354776, PHY-0354822 and PHY-0653342.


\appendix


\section{pQCD and NRQCD projection operators}\label{app:proj}

\subsection{Projection operators in pQCD factorization}

The heavy quark pair fragmentation function to a physical heavy quarkonium is defined in Eq.~(\ref{eq:QQFF}) \cite{KMQS-hq1}, in which there are two operators ${\cal P}_{ij,kl}^{(s)}(p_c)$ and ${\cal C}_{ab,cd}^{[I]}$\,, projecting the fragmenting heavy quark pair to a particular spin and color state.
Subscripts $i,j,k,l$ are the spinor indices and $a,b,c,d$ label the color of each field.
All of the definitions are given in \cite{KMQS-hq1}; we list them below for readers' convenience. Some explanations related to our calculation are also included.

The definitions of ${\cal P}_{ij,kl}^{(s)}(p_c)$ in $D$ dimensions are
\begin{linenomath*}
\begin{subequations}
\begin{align}
{\cal P}^{(v)}(p_c)_{ij,kl} & =\frac{1}{4p_c\cdot {\hat{n}}}\,(\gamma\cdot {\hat{n}})_{ij}\,\frac{1}{4p_c\cdot {\hat{n}}}\,(\gamma\cdot {\hat{n}})_{kl},\label{eq:pQCDv}\\
{\cal P}^{(a)}(p_c)_{ij,kl} & =\frac{1}{4p_c\cdot {\hat{n}}}\,\frac{\left[\gamma\cdot {\hat{n}}\,, \gamma_5\right]_{ij}}{2}\,\frac{1}{4p_c\cdot {\hat{n}}}\,\frac{\left[\gamma\cdot {\hat{n}}\,, \gamma_5\right]_{kl}}{2},\label{eq:pQCDa}\\
{\cal P}^{(t)}(p_c)_{ij,kl} & =\frac{1}{D-2}\sum_{\rho=1,2,\cdots,D-2}\frac{1}{4p_c\cdot {\hat{n}}}\,(\gamma\cdot {\hat{n}}\, \gamma^\rho_\bot)_{ij}\,\frac{1}{4p_c\cdot {\hat{n}}}\,(\gamma\cdot {\hat{n}}\,\gamma^\rho_\bot)_{kl}\label{eq:pQCDt},
\end{align}
\end{subequations}
\end{linenomath*}
where the superscripts $(v)$, $(a)$ or $(t)$ represent that the heavy quark pair is in a vector, axial-vector or tensor state, respectively, and ${\hat{n}}$ is a light-like vector, defined in Sec.~\ref{sec:pQCDFac}. To keep the charge conjugation invariance of the axial-vector heavy quark pair fragmentation function in dimensional regularization, 
we use $\left[\gamma\cdot {\hat{n}}\,, \gamma_5\right]/2=(\gamma\cdot {\hat{n}}\,\gamma_5-\gamma_5\,\gamma\cdot {\hat{n}})/2$ instead of $\gamma\cdot {\hat{n}}\, \gamma_5$.

The definitions of color projection operators ${\cal C}_{ab,cd}^{[I]}$ are
\begin{linenomath*}
\begin{subequations}
\begin{align}
{\cal C}^{[1]}_{ab,cd}&=\frac{1}{\sqrt{N_c}}\delta_{a,b}\frac{1}{\sqrt{N_c}}\delta_{c,d},\label{eq:pQCDc1}\\
{\cal C}^{[8]}_{ab,cd}&=\frac{2}{N_c^2-1}\sum_{f}(t^{(F)}_f)_{ab}\,(t^{(F)}_f)_{cd}\label{eq:pQCDc8}.
\end{align}
\end{subequations}
\end{linenomath*}

In Eqs.~(\ref{eq:QQFFQQ}), (\ref{eq:QQFFQQM2}) and (\ref{eq:QQFFQQA}), we split ${\cal P}^{(s)}(p_c)$ and ${\cal C}^{b}$ into products of several operators and normalization factors as
\begin{linenomath*}
\begin{subequations}
\label{eq:divprojection}
\begin{align}
{\cal P}^{(s)}&=\frac{\Gamma_s(p_c)\Gamma_{s}^\dag(p_c)}{N_s} P_s,\label{eq:spinorprojsplit}\\
{\cal C}^{b}&=\frac{C_b C_b^\dag}{N_b}\label{eq:colorprojsplit},
\end{align}
\end{subequations}
\end{linenomath*}
where the indices are suppressed, and all operators are understood to be inserted in the proper location as they are in Eqs.~(\ref{eq:QQFFQQ}) and (\ref{eq:QQFFQQM2}).  In Eq.~(\ref{eq:divprojection}), $s$ could be vector ($v$), axial-vector ($a$), or tensor ($t$), and $b$ could be ``$1$'' for color singlet or ``$8$'' for color octet. The operators in Eq.~(\ref{eq:spinorprojsplit}) are defined in $D$-dimension as
\begin{linenomath*}
\begin{subequations}
\begin{align}
\Gamma_v(p_c)&=\frac{\gamma\cdot {\hat{n}}}{4 p_c\cdot {\hat{n}}}
\text{\,\,\,,\,\,} N_v=1,
\\
\Gamma_a(p_c)&=\frac{[\gamma\cdot {\hat{n}}\, ,\gamma_5]}{8 p_c\cdot {\hat{n}}}
\text{\,\,\,,\,\,} N_a=1,
\\
\Gamma_t(p_c)&=\frac{\gamma\cdot {\hat{n}}\,\gamma^\rho}{4 p_c\cdot {\hat{n}}}
\text{\,\,\,,\,\,} N_t={D-2},
\end{align}
\end{subequations}
\end{linenomath*}
and
\begin{linenomath*}
\begin{subequations}
\begin{align}
P_v(p_c)&=P_a(p_c)=1,\\
P_t(p_c)&=-g_{\rho\rho'}+\frac{(p_c)_\rho {\hat{n}}_{\rho'}+(p_c)_{\rho'} {\hat{n}}_{\rho}}{p_c\cdot {\hat{n}}}-\frac{p_c^2}{(p_c\cdot {\hat{n}})^2}{\hat{n}}_\rho {\hat{n}}_{\rho'},
\end{align}
\end{subequations}
\end{linenomath*}
where Lorentz index $\rho'$ is the counterpart of $\rho$ in $\Gamma_t^\dag(p_c)$.

The color operators in Eq.~(\ref{eq:colorprojsplit}) are defined as
\begin{linenomath*}
\begin{subequations}\label{eq:c}
\begin{align}
C_{1}&=\frac{\mathbf{1}}{\sqrt{N_c}}
\text{\,\,\,,\,\,} N_{1}=1,
\\
C_{8}&=\sqrt{2}\,t^{(F)}_a
\text{\,\,\,,\,\,} N_{8}={N_c^2-1},
\end{align}
\end{subequations}
\end{linenomath*}
where $\mathbf{1}$ is a $3\times3$ unit matrix, the superscript $(F)$ represents the fundamental representation of SU(3), and subscript $a$ is summed between $C_8$ and $C_8^\dag$.

\subsection{Projection operators in NRQCD factorization}

NRQCD projectors $N^\text{NR}_{b'}$ in Eq.~\eqref{eq:QQFFQQ} and $C^\text{NR}_{b'}$ in Eq.~\eqref{eq:QQFFQQA} are the same as color projection operators of pQCD factorization in Eq.~(\ref{eq:c}), that is
\begin{linenomath*}
\begin{align}
C^\text{NR}_{b'}=C_{b'}\text{\,\,\,,\,\,} N^\text{NR}_{b'}=N_{b'}.
\end{align}
\end{linenomath*}
However, the meaning of $N_{8}=N_c^2-1$ and that of $N^\text{NR}_{8}=N_c^2-1$ are significantly different. The former indicates that the $\cc$-pair FFs are defined to average over the color of the fragmenting pair. In contrast, the later means that we need to average over the color of the NRQCD states in the short-distance coefficients, since the NRQCD LDMEs are defined to sum over all possible color states of the heavy quark pair.

The NRQCD spinor projection operators $\Gamma_i^{\text{NR}}$ in Eq.~\eqref{eq:QQFFQQA} are given by
\begin{linenomath*}
\begin{subequations}
\begin{align}
\Gamma^\text{NR}_{i}(p)=
&
\frac{1}{\sqrt{8m_Q^3}}(\frac{\slashed{p}}{2}-\slashed{q}_r-m_Q)\gamma_5(\frac{\slashed{p}}{2}+\slashed{q}_r+m_Q)
\qquad
\text{if $i$ is spin singlet},
\\
\Gamma^\text{NR}_{i}(p)=
&
\frac{1}{\sqrt{8m_Q^3}}(\frac{\slashed{p}}{2}-\slashed{q}_r-m_Q)\gamma^\beta(\frac{\slashed{p}}{2}+\slashed{q}_r+m_Q)
\qquad
\text{if $i$ is spin triplet},
\end{align}
\end{subequations}
\end{linenomath*}
where $m_Q$ is the heavy quark mass. The factor $1/({8m_Q^3})^{1/2}$ is partly caused by different normalizations between individual heavy quark and the pair, and we refer the interested readers to Ref.~\cite{Kang:2013hta}, for example, for a more detailed discussion.

The normalization factors $N^\text{NR}_i$ in Eq.~\eqref{eq:QQFFQQ} are defined as the number of states in $D$-dimension (number of color states are not counted here),
\begin{linenomath*}
\begin{subequations}
\begin{align}
&N^\text{NR}_{\statenc{1}{S}{0}}=N^\text{NR}_{\statenc{3}{P}{0}}=1,\\
&N^\text{NR}_{\statenc{3}{S}{1}}=N^\text{NR}_{\statenc{1}{P}{1}}=D-1,\\
&N^\text{NR}_{\statenc{3}{P}{1}}=\frac{1}{2}(D-1)(D-2),\\
&N^\text{NR}_{\statenc{3}{P}{2}}=\frac{1}{2}(D+1)(D-2),\\
&\sum_{J=0,1,2} N^\text{NR}_{\statenc{3}{P}{J}}=(D-1)^2.
\end{align}
\end{subequations}
\end{linenomath*}
$P^\text{NR}_{i}$ in Eq.~(\ref{eq:QQFFQQM2}) are defined as,
\begin{linenomath*}
\begin{subequations}
\begin{align}
&P^\text{NR}_{\statenc{1}{S}{0}}=1,\\
&P^\text{NR}_{\statenc{3}{S}{1}}=\mathbb{P}^{\beta\beta'}(p),\\
&P^\text{NR}_{\statenc{1}{P}{1}}=\mathbb{P}^{\alpha\alpha'}(p),\\
&P^\text{NR}_{\statenc{3}{P}{0}}=\frac{1}{D-1}\mathbb{P}^{\alpha\beta}(p)\mathbb{P}^{\alpha'\beta'}(p),\\
&P^\text{NR}_{\statenc{3}{P}{1}}=\frac{1}{2}(\mathbb{P}^{\alpha\alpha'}(p)\mathbb{P}^{\beta\beta'}(p)-\mathbb{P}^{\alpha\beta'}(p)\mathbb{P}^{\beta\alpha'}(p)),\label{eq:projP3p1}\\
&P^\text{NR}_{\statenc{3}{P}{2}}=\frac{1}{2}(\mathbb{P}^{\alpha\alpha'}(p)\mathbb{P}^{\beta\beta'}(p)+\mathbb{P}^{\alpha\beta'}(p)\mathbb{P}^{\beta\alpha'}(p))-\frac{1}{D-1}\mathbb{P}^{\alpha\beta}(p)\mathbb{P}^{\alpha'\beta'}(p),
\end{align}
\end{subequations}
\end{linenomath*}
where $\mathbb{P}^{\mu\nu}(p)$ is given by
\begin{linenomath*}
\begin{align}
\mathbb{P}^{\alpha\alpha'}(p)=-g^{\alpha\alpha'}+\frac{p^\alpha p^{\alpha'}}{p^2}\, ,
\end{align}
\end{linenomath*}
and Lorentz index $\alpha$ will be contracted with the Lorentz index from the derivative in Eq.~(\ref{eq:QQFFQQA}), and the primed Lorentz indices are for the complex conjugate of the amplitude, which are the counterparts of the unprimed ones in the amplitude.


\section{$\gamma_5$ in dimensional regularization}\label{app:gamma5}

The inconsistency between the following two properties of $\gamma_5$ in $D$-dimension
\begin{subequations}
\begin{align}
\text{Tr}[\gamma_5\gamma_\mu\gamma_\nu\gamma_\rho\gamma_\sigma]&=-4i\,\epsilon_{\mu\nu\rho\sigma},\label{eq:gamma5four}\\
\left[\gamma_5,\gamma_\alpha\right]&=0,\label{eq:gamma5com}
\end{align}
\end{subequations}
is well known~\cite{Breitenlohner:1977hr}.  Many $\gamma_5$ schemes in $D$-dimension have been proposed, such as 't Hooft-Veltman scheme~\cite{'tHooft:1972fi, Breitenlohner:1977hr} and Kreimer scheme~\cite{Kreimer:1989ke,Korner:1991sx,Kreimer:1993bh}. In principle, we can use any scheme as long as it is self-consistent. Although the resulted short-distance coefficients can differ by using different schemes, the difference is IR and UV finite at NLO calculation. Thus, one can perform a finite renormalization to relate results of different schemes. In our present work, we use a Kreimer-like scheme, while leave the finite renormalization and scheme independence discussion in future publication. In the Kreimer scheme, one needs to choose a ``reading point''.
As in our calculation all traces have zero, one or two $\gamma_5$'s, we adopt the following choice.

For traces with only one $\gamma_5$, we ``read'' or start the spinor trace in the amplitude from the $\gamma_5$, then use~\cite{West1993286}
\begin{align}
\text{Tr}[\gamma_5\gamma_{\alpha_1}\cdots\gamma_{\alpha_n}]=\frac{2}{n-4}\sum_{i=2}^{n}\sum_{j=1}^{i-1}(-1)^{i+j+1}g_{\alpha_i\alpha_j}\text{Tr}[\gamma_5\prod_{\substack{k=1,\\k\neq i,j}}^{n}\gamma_{\alpha_k}],
\end{align}
recursively until the trace involves only four $\gamma_\alpha$'s. For $n=4$, we use
\begin{align}
\text{Tr}[\gamma_5\gamma_\mu\gamma_\nu\gamma_\rho\gamma_\sigma] \text{Tr}[\gamma_5\gamma^\mu\gamma^\nu\gamma^\rho\gamma^\sigma]=16 D (D-1)(D-2)(D-3).
\end{align}
For traces with even number of $\gamma_5$'s, the reading point is in fact irrelevant, because we can always remove $\gamma_5$'s  by using Eq.~\eqref{eq:gamma5com} and
\begin{align}
\gamma_5\,\gamma_5=1.
\end{align}


\section{Single-Parton fragmentation functions}\label{app:SinglePartonFF}

In terms of the NRQCD factorization, the heavy quarkonium fragmentation functions from a single-parton are factorized in the form
\begin{align}
\begin{split}\label{eq:NRQCDFacSingle}
D_{f \to H}(z;m_Q,\mu_0)
=&
\sum_{[\cc(n)]}
\pi \as \Big\{\hat{d}^{\,(1)}_{f \to [\cc(n)]}(z;m_Q,\mu_0,\mu_\Lambda)\\
&+ \left(\frac{\as}{\pi}\right)\,\hat{d}^{\,(2)}_{f \to [\cc(n)]}(z;m_Q,\mu_0,\mu_\Lambda)+O(\alpha_s^2)\Big\}
\times
\frac{{\langle \mathcal{O}_{[\cc(n)]}^{H}(\mu_\Lambda)\rangle}}{m_Q^{2L+3}},
\end{split}
\end{align}
where $\mu_0$ (or $\mu_\Lambda$) is pQCD (or NRQCD) factorization scale,
$f$ could be gluon ($g$), light quark ($q$), charm quark ($c$), bottom quark ($b$), or their anti-particles,
$[\cc(n)]$ is an intermediate NRQCD $\cc$ state with quantum number $n=\state{{(2S+1)}}{L}{J}{1,8}$,
$H$ could be $\eta_c$, $J/\psi$, $\psip$, $h_c$, $\chi_{cJ}$, or their bottomonia counterparts,
and LDME $\langle \mathcal{O}_{\cc[n]}^{H}\rangle$ summarizes the nonperturbative physics
for the $[\cc(n)]$-pair to evolve into a heavy quarkonium $H$ at the energy scale below $\mu_\Lambda$.
The denominator $m_Q^{-(2L+3)}$ is introduced so that $\hat{d}^{\,(1)}$ and $\hat{d}^{\,(2)}$ are dimensionless.

Color singlet NRQCD LDMEs could be related to the value of (or the derivative of) heavy quarkonium wave functions at the origin, such as
\begin{align}
\begin{split}\label{eq:LDME1S01}
\langle \mathcal{O}_{[c\bar{c}(\CSaSz)]}^{\ \eta_c}\rangle
=
\frac{1}{4\pi}|R_{\eta_c}(0)|^2,
\end{split}
\\
\begin{split}\label{eq:LDME3S11}
\langle \mathcal{O}_{[c\bar{c}(\CScSa)]}^{\ J/\psi}\rangle
=
\frac{3}{4\pi}|R_{ J/\psi}(0)|^2,
\end{split}
\\
\begin{split}\label{eq:LDME1P11}
\langle \mathcal{O}_{[c\bar{c}(\CSaPa)]}^{\ h_c}\rangle
=
\frac{9}{4\pi}|R'_{ h_c}(0)|^2,
\end{split}
\\
\begin{split}\label{eq:LDME3PJ1}
\langle \mathcal{O}_{[c\bar{c}(\CScPj)]}^{\ \chi_{cJ}}\rangle
=
\frac{3(2J+1)}{4\pi}|R'_{ \chi_{cJ}}(0)|^2,
\end{split}
\end{align}
and similar relations are existed for LDMEs of producing bottomonia.
Values of these wave functions at the origin could be either calculated from potential model,
or fixed by data on heavy quakonium decay.
In contrast, color octet NRQCD LDMEs could only be extracted from data of heavy quakonium production at present.

In the rest of this appendix we list short-distance coefficients for all single-parton fragmentation functions  to $S$-wave and $P$-wave $\cc$-pair up to order $O(\alpha_s^2)$.
At ${\cal O}(\alpha_s)$, we have
\begin{align}\label{eq:gto3s180}
&\hat{d}^{\,(1)}_{g\to \COcSa}
=\frac{\delta(1-z)}{(3-2\epsilon)(N_c^2-1)},
\end{align}
while all other channels vanish. Results at $O(\alpha_s^2)$ are given in the following.


\subsection{Gluon Fragmentation Functions}
\begin{align}
\begin{split}\label{eq:gto1s01}
\hat{d}^{\,(2)}_{g\to \CSaSz}
=\frac{1}{N_c}
\big\{(1-z)\text{ln}[1-z]-z^2+\frac{3}{2}z\big\},
\end{split}\\
\begin{split}\label{eq:gto3s11}
\hat{d}^{\,(2)}_{g\to \CScSa }
=0,
\end{split}\\
\begin{split}\label{eq:gto1p11}
&\hat{d}^{\,(2)}_{g\to \CSaPa}
=0,
\end{split}\\
\begin{split}\label{eq:gto3pj1}
&\hat{d}^{\,(2)}_{g\to \CScPj}
=\frac{4}{9 N_c}
\Big\{
\Big[
\frac{Q_J}{2J+1}-\frac{1}{2}\text{ln}\big(\frac{\mu_\Lambda^2}{4m_Q^2}\big)
\Big]\delta(1-z)
+\frac{z}{(1-z)_+}
+\frac{P_J(z)}{2J+1}
\Big\},
\end{split}\\
\begin{split}\label{eq:gto3s18}
&\hat{d}^{\,(2)}_{g\to \COcSa}
=\frac{1}{12C_F}
\Big[
A(\mu_0)\delta(1-z)
+
\frac{1}{N_c}
P_{gg}(z)\Big(\text{ln}(\frac{\mu_0^2}{4m_Q^2})-1\Big)
\\
&\hspace{2cm}
+\frac{2(1-z)}{z}
-\frac{4(1-z+z^2)^2}{z}\left(\frac{\text{ln}(1-z)}{1-z}\right)_+
\Big],
\end{split}\\
\begin{split}\label{eq:gto1p18}
&\hat{d}^{\,(2)}_{g\to \COaPa}
=\frac{1}{12C_F}
\ [(1-z)\text{ln}(1-z)-z^2+\frac{3}{2}z],
\end{split}\\
\begin{split}\label{eq:gto1s08}
\hat{d}^{\,(2)}_{g\to \COaSz}
=\frac{B_F}{C_F}
\times
\hat{d}^{\,(2)}_{g\to \CSaSz}\, ,
\end{split}\\
\begin{split}\label{eq:gto3pj8}
\hat{d}^{\,(2)}_{g\to \COcPj}
=\frac{B_F}{C_F}
\times
\hat{d}^{\,(2)}_{g\to \CScPj}\, ,
\end{split}
\end{align}
where
\begin{align}
\begin{split}\label{eq:BF}
B_F
=
\frac{N_c^2-4}{4N_c},
\end{split}\\
\begin{split}
Q_0=\frac{1}{4},\hspace{2cm} Q_1=\frac{3}{8},\hspace{2cm} Q_2=\frac{7}{8},
\end{split}\\
\begin{split}
P_{0}(z)
=\frac{z(85-26z)}{8}+ \frac{9(5-3z)}{4}\text{ln}(1-z),
\end{split}\\
\begin{split}
P_{1}(z)
=-\frac{3 z(1+4z)}{4},
\end{split}\\
\begin{split}
P_{2}(z)
=\frac{5 z(11-4z)}{4}+ 9(2-z)\text{ln}(1-z),
\end{split}\\
\begin{split}
A(\mu)
=\frac{\beta_0}{N_c}\big[\text{ln}\big(\frac{\mu^2}{4m_Q^2}\big)+\frac{13}{3} \,\big]
+\frac{4}{N_c^2}-\frac{\pi^2}{3}+\frac{16}{3}\text{ln}2,
\end{split}\\
\begin{split}
P_{gg}(z)
=2N_c\Big[\frac{z}{(1-z)_+}+\frac{1-z}{z}+z(1-z)+\frac{\beta_0}{2N_c}\delta(1-z)\Big],
\end{split}\\
\begin{split}
\beta_0=\frac{11 N_c-2n_f}{6}.
\end{split}
\end{align}


\subsection{Same Flavor Heavy (Anti-)Quark Fragmentation Functions}

Heavy quark $Q$ has the same flavor as the outgoing $\cc$-pair.
\begin{align}
\begin{split}\label{eq:Qto1s01}
\hat{d}^{\,(2)}_{Q \to \CSaSz}
=\frac{2}{3}
\frac{C_F^2}{N_c}
\frac{(z-1)^2}{(z-2)^6}
\ z(3 z^4-8 z^3+8 z^2+48),
\end{split}\\
\begin{split}\label{eq:Qto3s11}
\hat{d}^{\,(2)}_{Q\to \CScSa}
=\frac{2}{3}
\frac{C_F^2}{N_c}
\frac{(z-1)^2}{(z-2)^6}
\ z(5 z^4-32 z^3+72 z^2-32 z+16),
\end{split}\\
\begin{split}\label{eq:Qto1p11}
&\hat{d}^{\,(2)}_{Q\to \CSaPa}
=\frac{2}{3}
\frac{C_F^2}{N_c}
\frac{(z-1)^2}{(z-2)^8}
\ z(9 z^6-56 z^5+140 z^4-160 z^3+176 z^2
-128 z+64),
\end{split}\\
\begin{split}\label{eq:Qto3p01}
&\hat{d}^{\,(2)}_{Q\to \CScPz}
=\frac{2}{9}
\frac{C_F^2}{N_c}
\frac{(z-1)^2}{(z-2)^8}
\ z(59 z^6-376 z^5+1060 z^4-1376 z^3+528 z^2
\\
&\hspace{2cm}
+384 z+192),
\end{split}\\
\begin{split}\label{eq:Qto3p11}
&\hat{d}^{\,(2)}_{Q\to \CScPa}
=\frac{8}{9}
\frac{C_F^2}{N_c}
\frac{(z-1)^2}{(z-2)^8}
\ z(7 z^6-54 z^5+202 z^4-408 z^3+496 z^2
-288 z+96),
\end{split}\\
\begin{split}\label{eq:Qto3p21}
&\hat{d}^{\,(2)}_{Q\to \CScPb}
=\frac{16}{45}
\frac{C_F^2}{N_c}
\frac{(z-1)^2}{(z-2)^8}
\ z(23 z^6-184 z^5+541 z^4-668 z^3+480 z^2\\
&\hspace{2cm}
-192 z+48),
\end{split}\\
\begin{split}\label{eq:Qto3s18}
&\hat{d}^{\,(2)}_{Q\to \COcSa}
=\frac{1}{12}
\frac{1}{N_c^3}
\frac{1}{(z-2)^6\,z}
\Big\{
N_c^2 (z^2-2 z+2) (z-2)^6 \,\text{ln} \big(\frac{\mu_0^2}{(z-2)^2 m_Q^2} \big)
\\
&
\hspace{1cm}
-N_c^2\, (z-2)^4 z^2 (z^2-10z+10)
-16\, N_c\, z (z-2)^2 (z^4-7 z^3+12 z^2-8 z+2)
\\
&
\hspace{1cm}
+2 (z-1)^2 z^2 (5 z^4-32 z^3+72 z^2-32 z+16)
\Big\},
\end{split}\\
\begin{split}\label{eq:Qto1s08}
\hat{d}^{\,(2)}_{Q\to \COaSz}
=\frac{1}{(N_c^2-1)^2}\times \hat{d}^{\,(2)}_{Q\to \CSaSz}\, ,
\end{split}\\
\begin{split}\label{eq:Qto1p18}
\hat{d}^{\,(2)}_{Q\to \COaPa}
=\frac{1}{(N_c^2-1)^2}\times \hat{d}^{\,(2)}_{Q\to \COaPa}\, ,
\end{split}\\
\begin{split}\label{eq:Qto3p08}
\hat{d}^{\,(2)}_{Q\to \COcPj}
=\frac{1}{(N_c^2-1)^2}\times \hat{d}^{\,(2)}_{Q\to \CScPj}\, ,
\end{split}\\
\begin{split}
\hat{d}^{\,(2)}_{\bar{Q}\to n}
=\hat{d}^{\,(2)}_{Q\to n},
\hspace{1cm}
\text{for any $n=\state{{2S+1}}{L}{J}{1,8}$}.
\end{split}
\end{align}

\subsection{Light (Anti-)Quark Fragmentation Functions}

Light quark $q$ could be $u$, $d$ or $s$.
\begin{align}\label{eq:qto3s18}
\begin{split}
\hat{d}^{\,(2)}_{q\to \COcSa}
=\frac{1}{12 N_c}\frac{1}{z}
\Big\{ (z^2 - 2 z + 2) \text{ln}\Big[\frac{\mu_0^2}{4m_Q^2(1-z)}\Big]-2 z^2\Big\},
\end{split}\\
\begin{split}
\hat{d}^{\,(2)}_{q\to n}
=0,
\hspace{1.65cm}
\text{for $n \neq \COcSa$},
\end{split}\\
\begin{split}
\hat{d}^{\,(2)}_{\bar{q}\to n}
=\hat{d}^{\,(2)}_{q\to n},
\hspace{1cm}
\text{for any $n=\state{{2S+1}}{L}{J}{1,8}$}.
\end{split}
\end{align}

\subsection{Different Flavor Heavy (Anti-)Quark Fragmentation Functions}

Heavy quark $Q'$ has a different flavor with outgoing $\cc$-pair.
\begin{align}
\begin{split}\label{eq:bto3s18}
\hat{d}^{\,(2)}_{Q'\to \COcSa}
=\frac{1}{12N_c}\frac{1}{z}
\left\{ (z^2 - 2 z + 2) \text{ln}\Big[\frac{\mu_0^2}{4m_Q^2(1-z+\frac{z^2\eta}{4})}\Big]- 2z^2\Big(1+\frac{1-z-\frac{z^2}{2}}{4-4z+z^2\eta}\eta\Big) \right\},
\end{split}\\
\begin{split}
\hat{d}^{\,(2)}_{Q'\to n}
=0,
\hspace{1.65cm}
\text{for $n \neq \COcSa$},
\end{split}\\
\begin{split}
\hat{d}^{\,(2)}_{\bar{Q}'\to n}
=\hat{d}^{\,(2)}_{Q'\to n},
\hspace{1cm}
\text{for any $n=\state{{2S+1}}{L}{J}{1,8}$},
\end{split}
\end{align}
where $\eta=m_{Q'}^2/m_Q^2$, {with $m_Q$ the mass of the heavy quark in the outgoing $\cc$-pair}.

\subsection{Comparison with Previous Results}
Many of the above results have been calculated and are available in the literature.  We present here a brief comparison with previous results.

For gluon fragmentation into a heavy quark pair, Eq.~(\ref{eq:gto1s01}) and Eq.~(\ref{eq:gto1s08}) confirm the results in Refs.~\cite{Braaten:1993rw} and \cite{Hao:2009fa}, respectively.
Eq.~(\ref{eq:gto3pj1}) verifies the result of Ref.~\cite{Braaten:1996rp} using the dimensional regularization, which is consistent with the earlier work in Ref.~\cite{Braaten:1994kd} evaluated in a cutoff regularization scheme.  Summing over all $J$, Eq.~(\ref{eq:gto3pj1}) is also consistent with the result in Ref.~\cite{Bodwin:2012xc}.  Eq.~(\ref{eq:gto3s18}) seems to have a very minor difference from the previous calculation of $g\to [\cc(\COcSa)]+X$ fragmentation \cite{Braaten:2000pc}. The minor difference seems to be caused by the derivation of $I_{ACD}$ in Eq. (A.11) of Ref.~\cite{Braaten:2000pc}. Our result for $I_{ACD}$ can be obtained by replacing $-6\ln^2 2$ in Eq. (A.11) of Ref.~\cite{Braaten:2000pc} with $-2\ln^2 2$.

For light-quark fragmentation into a $\cc$ pair, Eq.~(\ref{eq:qto3s18}) confirms the result in Ref.~\cite{Ma:1995vi}.

For heavy quark fragmentation into a $\cc$ pair, Eqs.~(\ref{eq:Qto1s01}) and (\ref{eq:Qto3s11}) confirm the results in Ref.~\cite{Braaten:1993mp}.
Eqs. (\ref{eq:Qto1p11})-(\ref{eq:Qto3p21}) and Eq. (\ref{eq:Qto1s08}) are the same as the results in Ref.~\cite{Yuan:1994hn}. But, our result in Eq.~(\ref{eq:Qto3s18}) is slightly different from both the result in Ref.~\cite{Yuan:1994hn} and that in Ref.~\cite{Ma:1995vi}, while the results from these two authors are slightly different from each other for this $Q\to [\cc(\COcSa)]+Q$ channel.


\section{Results for double-parton FF's}\label{app:doubleresults}

\subsection{Definitions and Notations}

Similar to Eq.~(\ref{eq:NRQCDFacSingle}), in terms of NRQCD factorization, the $\cc$ pair fragmentation functions could be factorized as
\begin{align}
\begin{split}\label{eq:NRQCDFacDouble}
{\mathcal D}_{[\cc(\kappa)]\to H}(z,\zeta_1,\zeta_2,\mu_0;m_Q)
&=
\sum_{[\cc(n)]}
\Big\{\hat{d}^{\,(0)}_{[\cc(\kappa)]\to [\cc(n)]}(z,\zeta_1,\zeta_2,\mu_0;m_Q,\mu_\Lambda)
\\
&\hspace{-3cm}+
\left(\frac{\as}{\pi}\right)
\hat{d}^{\,(1)}_{[\cc(\kappa)]\to [\cc(n)]}(z,\zeta_1,\zeta_2,\mu_0;m_Q,\mu_\Lambda)
+O(\alpha_s^2)\Big\}
\times
\frac{\langle \mathcal{O}_{[\cc(n)]}^{H}(\mu_\Lambda)\rangle}{m_Q^{2L+1}},
\end{split}
\end{align}
where $[\cc(\kappa)]$ is a perturbatively produced fragmenting heavy quark pair in a particular spin and color state $\kappa$, which could be vector ($v$), axial-vector ($a$) or tensor ($t$), with either color singlet or octet.
Again, the denominator $m_Q^{-(2L+1)}$ is used so that $\hat{d}^{\,(0)}$ and $\hat{d}^{\,(1)}$ are dimensionless.

In the rest of this appendix we list our results of the short-distance coefficients for all $\cc$-pair fragmentation functions into S-wave NRQCD $\cc$-pair up to NLO.  In the following, we omit the subscript $\cc$ to use the notation, $\hat{d}^{\,(j)}_{\,\kappa \to n}\ (j=0,1)$ instead of
$\hat{d}^{\,(j)}_{[\cc(\kappa)]\to [\cc(n)]}(z,\zeta_1,\zeta_2,\mu_0;m_Q,\mu_\Lambda)$.
Note that we do not list any results that vanish except $\hat{d}^{\,(1)}_{\tone \to \COaSz}$,
which is equal to zero only in our present $\gamma_5$ scheme.

\subsection{LO results}
\begin{align}
\begin{split}\label{eq:v1to3s11LO}
\hat{d}^{\,(0)}_{\vone \to \CScSa}
=\frac{1}{2(3-2\epsilon)}\delta(\zeta_1)\delta(\zeta_2)\delta(1-z),
\end{split}\\
\begin{split}\label{eq:a1to1s01LO}
\hat{d}^{\,(0)}_{\aone \to \CSaSz}
=\frac{1}{2}\delta(\zeta_1)\delta(\zeta_2)\delta(1-z),
\end{split}\\
\begin{split}\label{eq:t1to3s11LO}
\hat{d}^{\,(0)}_{\tone \to \CScSa}
=\frac{1}{2(3-2\epsilon)}\delta(\zeta_1)\delta(\zeta_2)\delta(1-z),
\end{split}\\
\begin{split}\label{eq:s18tos28LO}
\hat{d}^{\,(0)}_{s^{[8]}\to \state{{2S+1}}{L}{J}{8}}
=\frac{1}{N_c^2-1}\hat{d}^{\,(0)}_{s^{[1]}\to \state{{2S+1}}{L}{J}{1}}(\zeta_1,\zeta_2,z),
\end{split}
\end{align}
where $s$ could be $v$, $a$ or $t$, and $\epsilon = (D-4)/2$.

\subsection{NLO results}

\begin{align}
\begin{split}\label{eq:v1to3s11NLO}
&\hat{d}^{\,(1)}_{\vone \to \CScSa}
=
\frac{1}{12}
C_F\delta(1-z)
\Big\{
\frac{3}{4}\, \DeltaZero\times \LogUV
+
\tilde{V}_{va}(\zeta_1,\zeta_2)
\,
\big( \LogUV -\frac{2}{3}\, \big)
\\
&\hspace{2cm}
+V_{1}(\zeta_1,\zeta_2)
\Big\}
,
\end{split}\\
\begin{split}\label{eq:v1to1s08NLO}
\hat{d}^{\,(1)}_{\vone \to \COaSz}
=
\frac{1}{8}
\frac{C_F}{(N_c^2-1)}
\DPlusOne
\,z (1-z)
\Big\{
\LogUV-2\,\text{ln}(2-2z)-3
\Big\}
,
\end{split}\\
\begin{split}\label{eq:v1to3s18NLO}
\hat{d}^{\,(1)}_{\vone \to \COcSa}
=\frac{1}{24}
\frac{C_F}{(N_c^2-1)}
\DMinusOne\,
\frac{z}{(1-z)}\,
\Big\{
(\LogUV-\frac{2}{3})
-2\,\text{ln}(2-2z)+2z^2-4z+\frac{5}{3}
\Big\}
,
\end{split}\\
\begin{split}\label{eq:v8to1s08NLO}
\hat{d}^{\,(1)}_{\veight \to \COaSz}
=\frac{1}{8}\frac{C_F}{(N_c^2-1)^2}\,
{\DMinusEight}\,
z (1-z)
\Big\{
\LogUV-2\,\text{ln}(2-2z)-3
\Big\}
,
\end{split}\\
\begin{split}\label{eq:v8to3s18NLO}
&
\hat{d}^{\,(1)}_{\veight \to \COcSa}
=
-\frac{z}{12}
\frac{C_F}{(N_c^2-1)^2}
\Big\{
\delta(1-z)
\Big[\,
\frac{3}{4}\DeltaZero \Big(\tilde{c} \times \LogUV+c_0\Big)
\\
&\hspace{1.7cm}
+\big(\tilde{V}_{va}(\zeta_1,\zeta_2)+\frac{N_c}{2}\,\tilde{V}_{g}(\zeta_1,\zeta_2)\big)\big( \LogUV-\frac{2}{3}\,\big)
+V_{1}(\zeta_1,\zeta_2)+\frac{N_c}{2}\,V_g(\zeta_1,\zeta_2)\Big]
\\
&\hspace{1.7cm}
-
\frac{\DPlusEight}{2\,(1-z)_+}
\big( \LogUV-\frac{2}{3}\,\big)
+
\DPlusEight\,R_1(z)
\Big\}
,
\end{split}\\
\begin{split}\label{eq:a1to1s01NLO}
&\hat{d}^{\,(1)}_{\aone \to \CSaSz}
=
\frac{1}{4}
C_F
\delta(1-z)
\Big\{
\big[\,\frac{3}{4}\,\DeltaZero +\tilde{V}_{va}(\zeta_1,\zeta_2)\big] \LogUV
+V_{2}(\zeta_1,\zeta_2)
\Big\}
,
\end{split}\\
\begin{split}\label{eq:a1to1s08NLO}
\hat{d}^{\,(1)}_{\aone \to \COaSz}
=
\frac{1}{8}
\frac{C_F}{(N_c^2-1)}
\DMinusOne
\frac{z}{ (1-z)}
\Big\{
\LogUV-2\,\text{ln}(2-2z)-1
\Big\}
,
\end{split}\\
\begin{split}\label{eq:a1to3s18NLO}
\hat{d}^{\,(1)}_{\aone \to \COcSa}
=
\frac{1}{24}
\frac{C_F}{(N_c^2-1)}
\DPlusOne
\,{z}{ (1-z)}
\Big\{
(\LogUV-\frac{2}{3})
-2\,\text{ln}(2-2z)-\frac{1}{3}
\Big\}
,
\end{split}\\
\begin{split}\label{eq:a8to1s08NLO}
&
\hat{d}^{\,(1)}_{\aeight \to \COaSz}
=
-\frac{z}{4}
\frac{C_F}{(N_c^2-1)^2}
\Big\{
\delta(1-z)\Big[\,
\frac{3}{4}\DeltaZero  \Big( \tilde{c} \times\LogUV +c_0\Big)
+\,\tilde{V}_{va}(\zeta_1,\zeta_2)\LogUV
\\
&\hspace{2cm}
+V_2(\zeta_1,\zeta_2)\Big]
-
\frac{\DPlusEight}{2\,(1-z)_+} \LogUV
+
\DPlusEight\,R_{2}(z)
\Big\}
,
\end{split}\\
\begin{split}\label{eq:a8to3s18NLO}
\hat{d}^{\,(1)}_{\aeight \to \COcSa}
=
\frac{1}{24}
\frac{C_F}{(N_c^2-1)^2}
\DMinusEight
\,{z}{ (1-z)}
\Big\{
(\LogUV-\frac{2}{3})
-2\,\text{ln}(2-2z)-\frac{1}{3}
\Big\}
,
\end{split}\\
\begin{split}\label{eq:t1to3s11NLO}
&\hat{d}^{\,(1)}_{\tone \to \CScSa}
=
\frac{1}{12}C_F
\delta(1-z)
\Big\{
\frac{3}{4}\, \DeltaZero \times \LogUV
+
\tilde{V}_t(\zeta_1,\zeta_2)
\,
\big( \LogUV -\frac{2}{3}\, \big)
\\
&\hspace{2cm}
+V_{3}(\zeta_1,\zeta_2)
\Big\}
,
\end{split}\\
\begin{split}\label{eq:t1to1s08NLO}
\hat{d}^{\,(1)}_{\tone \to \COaSz}
=0,
\end{split}\\
\begin{split}\label{eq:t1to3s18NLO}
\hat{d}^{\,(1)}_{\tone \to \COcSa}
=
\frac{1}{24}
\frac{C_F}{(N_c^2-1)}
\DMinusOne\,
\frac{z(z^2-2z+2)}{(1-z)}
\Big\{
(\LogUV-\frac{2}{3})
-2\,\text{ln}(2-2z)-\frac{1}{3}
\Big\}
,
\end{split}\\
\begin{split}\label{eq:t8to3s18NLO}
&
\hat{d}^{\,(1)}_{\teight \to \COcSa}
=
-\frac{z}{12}
\frac{C_F}{(N_c^2-1)^2}
\Big\{
\delta(1-z)
\Big[\,
\frac{3}{4}\DeltaZero \Big(\tilde{c} \times \LogUV+c_0\Big)
\\
&\hspace{1.8cm}
+(\tilde{V}_t(\zeta_1,\zeta_2)+\frac{N_c}{2}\tilde{V}_{tg}(\zeta_1,\zeta_2))\big( \LogUV-\frac{2}{3}\,\big)
+V_{3}(\zeta_1,\zeta_2)+\frac{N_c}{2}{V}_{tg}(\zeta_1,\zeta_2)\Big]
\\
&\hspace{1.8cm}
-
\DPlusEight \, \frac{(z^2-2z+2)}{2\,(1-z)_+} \big( \LogUV-\frac{2}{3}\,\big)
+
\DPlusEight\,R_3(z)
\Big\}
,
\end{split}\\
\begin{split}\label{eq:v8to1s01NLO}
\hat{d}^{\,(1)}_{s^{[8]} \to \state{{2S+1}}{L}{J}{1}}
=\hat{d}^{\,(1)}_{s^{[1]} \to \state{{2S+1}}{L}{J}{8}},
\end{split}
\end{align}
where the zero result in Eq.~(\ref{eq:t1to1s08NLO}) depends on the $\gamma^5$ scheme,
the $2/3$ in factor $(\text{ln}[\mu_0^2/m_Q^2]-2/3)$ comes from the $\epsilon$-dependence of LO results, $s$ could be $v$, $a$ or $t$. $\tilde{V}$, $V$, $R$ and $c$ are defined as
%
\begin{align}
\begin{split}
\tilde{V}_{va}(\zeta_1,\zeta_2)
=
\delta(\zeta_2)\Big\{
\Plusa-\frac{1}{2}\Plusz{\zeta_1+1}
\Big\}
+(\zeta_1 \leftrightarrow \zeta_2)
,
\end{split}\\
\begin{split}
\tilde{V}_{g}(\zeta_1,\zeta_2)
=
\delta(\zeta_2)\Big\{\Plusz{1-\zeta_1^2} \Big\}
+(\zeta_1 \leftrightarrow \zeta_2),
\end{split}\\
\begin{split}
\tilde{V}_{t}(\zeta_1,\zeta_2)
=
\delta(\zeta_2)\Big\{\Plusa-\Plusz{1} \Big\}
+(\zeta_1 \leftrightarrow \zeta_2),
\end{split}\\
\begin{split}
\tilde{V}_{tg}(\zeta_1,\zeta_2)
=
\delta(\zeta_2)\ \Plusz{1}
+(\zeta_1 \leftrightarrow \zeta_2),
\end{split}\\
\begin{split}
&V_{1}(\zeta_1,\zeta_2)
=
\delta(\zeta_2)
\Big\{
\Plusb
-\PlusLoga-\frac{1}{3}\Plusa+\frac{1}{2}\Plusz{\zeta_1\text{ln}(\zeta_1^2)}
\\
&\hspace{2cm}
+\frac{7}{6}\Plusz{\zeta_1}
+\frac{1}{2}\,\Plusz{\text{ln}(\zeta_1^2)}-\frac{11}{6}\Plusz{1}
\Big\}
+(\zeta_1 \leftrightarrow \zeta_2)
,
\end{split}\\
\begin{split}
&V_{2}(\zeta_1,\zeta_2)
=
\delta(\zeta_2)\Big\{
\Plusb
-\PlusLoga-\Plusa+\frac{1}{2}\Plusz{\zeta_1\text{ln}(\zeta_1^2)}
\\
&\hspace{2cm}
+\frac{1}{2}\Plusz{\zeta_1}
+\frac{1}{2}\,\Plusz{\text{ln}(\zeta_1^2)}-\frac{1}{2}\Plusz{1}
\Big\}
+(\zeta_1\leftrightarrow \zeta_2)
,
\end{split}\\
\begin{split}
&V_{3}(\zeta_1,\zeta_2)
=
\delta(\zeta_2)\Big\{
\Plusb
-\PlusLoga-\frac{1}{3}\Plusa
+\Plusz{\text{ln}(\zeta_1^2)}
-\frac{2}{3}\Plusz{1}
\Big\}
\\
&\hspace{2cm}
+(\zeta_1 \leftrightarrow \zeta_2),
\end{split}\\
\begin{split}
&V_{g}(\zeta_1,\zeta_2)
=
\delta(\zeta_2)\Big\{
\Plusz{\zeta_1^2\mylog{\zeta_1^2}}
-\Plusz{\mylog{\zeta_1^2}}
+\frac{2}{3}\Plusz{1-\zeta_1^2}
\Big\}
+(\zeta_1 \leftrightarrow \zeta_2),
\end{split}\\
\begin{split}
&V_{tg}(\zeta_1,\zeta_2)
=
\delta(\zeta_2)\Big\{
\frac{2}{3}\Plusz{1}
-\Plusz{\mylog{\zeta_1^2}}
\Big\}
+(\zeta_1 \leftrightarrow \zeta_2),
\end{split}\\
\begin{split}
&
R_1(z)
=
\left(\frac{\text{ln}(2-2z)}{1-z}\right)_+
+\frac{1}{6}\frac{1}{(1-z)_+}
-(1-z)
\, ,
\end{split}\\
\begin{split}
&
R_2(z)
=
\left(\frac{\text{ln}(2-2z)}{1-z}\right)_+
+\frac{1}{2}\frac{1}{(1-z)_+}
\, ,
\end{split}\\
\begin{split}
&
R_{3}(z)
=
(z^2-2z+2)
\Big\{
\left(\frac{\text{ln}(2-2z)}{1-z}\right)_+ +\frac{1}{6}\frac{1}{(1-z)_+}
\Big\},
\end{split}\\
\begin{split}
\tilde{c}
=
1-N_c^2(1+\frac{4}{3}\logtwo),
\end{split}\\
\begin{split}
c_0
=
\frac{4}{3}\,N_c^2\big[(\logtwo)^2 +\logtwo -1\big].
\end{split}
\end{align}
Definitions of the plus-distributions are given in Eq.~(\ref{eq:zplus}) and Eq.~(\ref{eq:plusminusalldef}).
The $\Delta$-functions are defined as
%
\begin{align}
\begin{split}\label{eq:Delta0}
\DeltaZero
=
4\,\delta(\zeta_1)\delta(\zeta_2),
\end{split}\\
\begin{split}\label{eq:Delta1}
\DPMOne
=4[\DAA \pm \DBB][\DXX \pm \DYY],
\end{split}\\
\begin{split}\label{eq:Delta8}
&\DPMEight
=4\big\{(N_c^2-2)[\DAA\DXX+\DBB\DYY]\\
&\hspace{1cm}
\mp 2[\DAA\DYY+\DBB\DXX] \big\}.
\end{split}
\end{align}

\subsection{Comparison with Other Calculations}
A similar calculation for the color singlet process $[\cc(\aone)] \to [\cc(\CSaSz)]$, in the terminology of distribution amplitude, was completed by two groups previously \cite{Bell:2008er,Ma:2006hc}, but, their results disagree with each other.  Our result in Eq.~(\ref{eq:a1to1s01NLO}) confirms the calculation of Ref.~\cite{Bell:2008er}.  For process $[\cc(\vone)] \to [\cc(\CScSa)]$, our result in Eq.~(\ref{eq:v1to3s11NLO}) disagree with the result obtained in \cite{Ma:2006hc}. Finally, we note that, soon after our paper was submitted, an independent calculation for $[\cc(\aone)] \to [\cc(\CSaSz)]$, $[\cc(\vone)] \to [\cc(\CScSa)]$ and $[\cc(\tone)] \to [\cc(\CScSa)]$ was also reported in Ref. \cite{Wang:2013ywc} in the terminology of distribution amplitude. Our results for these three channels agree with that calculated in Ref. \cite{Wang:2013ywc}.


\providecommand{\href}[2]{#2}\begingroup\raggedright\endgroup

\end{document}